\documentclass[preprint2, trackchanges, twocolappendix]{aastex701}
\usepackage{amsmath}
\usepackage[export]{adjustbox}

\nolinenumbers

\received{13 October 2025}
\accepted{24 November 2025} 

\begin{document}

\title{Feeding the Dead: Neutral Gas Inflow in a Long-Quenched Ancient Massive Galaxy at $z\sim 2.7$ Observed with JWST/NIRSpec}

\correspondingauthor{Davide Bevacqua}

\author[orcid=0000-0001-8863-2472,sname='Bevacqua']{Davide Bevacqua}
\affiliation{Physics and Astronomy Department, Tufts University, 574 Boston Avenue, Medford, MA 02155, USA}
\affiliation{INAF - Osservatorio Astronomico di Brera, via Brera 28, 20121 Milano, Italy}
\email[show]{davide.bevacqua@inaf.it}

\author[sname='Marchesini']{Danilo Marchesini} 
\affiliation{Physics and Astronomy Department, Tufts University, 574 Boston Avenue, Medford, MA 02155, USA}
\email{danilo.marchesini@tufts.edu}

\author[sname='Saracco']{Paolo Saracco} 
\affiliation{INAF - Osservatorio Astronomico di Brera, via Brera 28, 20121 Milano, Italy}5
\email{paolo.saracco@inaf.it}

\author[sname='La Barbera']{Francesco La Barbera} 
\affiliation{INAF - Osservatorio Astronomico di Capodimonte, Via Moiariello 16, 80131, Naples, Italy}
\email{francesco.labarbera@inaf.it}

\author[sname='Pan']{Richard Pan}
\affiliation{Physics and Astronomy Department, Tufts University, 574 Boston Avenue, Medford, MA 02155, USA}
\email{danilo.marchesini@tufts.edu}

\author[orcid=0000-0002-5615-6018, sname='Belli']{Sirio Belli}
\affiliation{Dipartimento di Fisica e Astronomia, Universit\'{a} di Bologna, Via Gobetti 93/2, I-40129, Bologna, Italy}
\email{sirio.belli@unibo.it}

\author[sname='Brammer']{Gabriel Brammer}
\affiliation{Niels Bohr Institute, University of Copenhagen, Jagtvej 128, Copenhagen, Denmark}
\email{gabriel.brammer@nbi.ku.dk}

\author[sname='De Marchi']{Guido De Marchi}
\affiliation{European Space Research and Technology Centre, Keplerlaan 1, 2200 AG Noordwijk, Netherlands}
\email{gdemarchi@cosmos.esa.int}

\author[sname='Ditrani']{Fabio R. Ditrani}
\affiliation{Universit\'{a} degli studi di Milano-Bicocca, Piazza della scienza, I-20125 Milano, Italy}
\affiliation{INAF-Osservatorio Astronomico di Brera, via Brera 28, I-20121 Milano, Italy}
\email{f.ditrani1@campus.unimib.it}

\author[sname='Giardino']{Giovanna Giardino}
\affiliation{ATG Europe for the European Space Agency, European Space Research and Technology Centre, Noordwijk, Netherlands}
\email{giovanna.giardino@esa.int}

\author[sname='Glazebrook']{Karl Glazebrook}
\affiliation{Centre for Astrophysics and Supercomputing, Swinburne University of Technology, PO Box 218, Hawthorn, VIC 3122, Australia}
\email{kglazebrook@swin.edu.au}

\author[sname='La Torre']{Valentina La Torre}
\affiliation{Physics and Astronomy Department, Tufts University, 574 Boston Avenue, Medford, MA 02155, USA}
\email{valentina.la_torre@tufts.edu}

\author[sname='Lin']{Jamie Lin}
\affiliation{Physics and Astronomy Department, Tufts University, 574 Boston Avenue, Medford, MA 02155, USA}
\email{jamie.lin@tufts.edu}

\author[sname='Muzzin']{Adam Muzzin}
\affiliation{Department of Physics and Astronomy, York University, 4700 Keele Street, Toronto, ON, M3J 1P3, Canada}
\email{muzzin@yorku.ca}

 \author[orcid=0000-0002-4430-8846, sname='Roy']{Namrata Roy} 
\affiliation{School of Earth and Space Exploration, Arizona State University, Tempe, AZ 85281, USA}
\affiliation{Department of Physics \& Astronomy, Johns Hopkins University, Baltimore, MD 21218, USA}
\email{namratar@asu.edu} 

\author[orcid=0000-0002-9334-8705, sname='Santini']{Paola Santini}
\affiliation{INAF - Osservatorio Astronomico di Roma, via di Frascati 33, 00078 Monte Porzio Catone, Roma Italy}
\email{paola.santini@inaf.it}

\author[orcid=0000-0003-0980-1499, sname='Vulcani']{Benedetta Vulcani}
\affiliation{INAF- Osservatorio astronomico di Padova, Vicolo Osservatorio 5, I-35122 Padova, Italy}
\email{benedetta.vulcani@inaf.it}

\author[orcid=0000-0003-3108-0624, sname='Watson']{Peter J. Watson}
\affiliation{INAF- Osservatorio astronomico di Padova, Vicolo Osservatorio 5, I-35122 Padova, Italy}
\email{peter.watson@inaf.it}

\author[orcid=0000-0002-9373-3865]{Xin Wang}
\affiliation{School of Astronomy and Space Science, University of Chinese Academy of Sciences (UCAS), Beijing 100049, China}
\affiliation{National Astronomical Observatories, Chinese Academy of Sciences, Beijing 100101, China}
\affiliation{Institute for Frontiers in Astronomy and Astrophysics, Beijing Normal University, Beijing 102206, China}
\email{xwang@ucas.ac.cn}

\begin{abstract}

We report the spectroscopic detection of neutral gas inflow into a massive ($M_* \simeq 4\times 10^{10} M_\odot$) quiescent galaxy observed at $z_{\rm{spec}} = 2.6576$ with JWST. From the redshifted absorption of the NaI doublet at $\lambda \lambda \,5890, 5896 \, \AA$, we estimate an inflow velocity $v=278^{+79}_{-79}$ km s$^{-1}$ and a column density $\log(N_{NaI}/\rm{cm^2}) = 13.02^{+0.03}_{-0.03}$. We derive the inflowing mass of the gas $M_{in} = 1.6^{+0.1}_{-0.1} \times 10^8 M_\odot$ and rate $\dot{M}_{in} = 19^{+6}_{-7} \, M_\odot \, \rm{yr}^{-1}$. The presence of several surrounding galaxies suggests that the galaxy may be accreting gas from nearby companions. However, we cannot confirm it with current data and the intergalactic medium or cosmic filaments are also viable sources of the inflowing gas. Despite the ongoing inflow, the galaxy remains quiescent, with an upper limit to the star formation rate of $0.2 \, M_\odot \, \rm{yr}^{-1}$. Moreover, its star formation history suggests that the galaxy has remained quiescent during the past $\sim1$ Gyr, with half of its stars formed by redshift $z_{50}=11^{+18}_{-3}$. We discuss that the inflow is not massive, dense, or long-lived enough to ignite significant star formation, or it is fueling low-level AGN activity instead. This is direct evidence that quiescent galaxies can accrete cold gas after their quenching while keeping their star formation subdued. Follow-up observations with JWST and ALMA will be needed to constraint the nature of the inflowing gas.

\end{abstract}



\section{Introduction} 


In the standard paradigm of galaxy formation, galaxies grow through the accretion of gas from the cosmic web, which fuels star formation (SF) and the growth of central supermassive black holes \citep[e.g.,][]{WhiteRees78, Dekel+09}. Over cosmic time, galaxies can run out of cold gas, stop their SF (or `quench'), and become quiescent. Despite the cessation of SF, quiescent galaxies can continue to interact with their surroundings, accreting gas from the cosmic web or close companions, eventually igniting new SF or active galactic nucleus (AGN) activity. 

Many observational studies \citep[e.g., ][]{Chauke+19, Mancini+19, Woodrum+22, Tacchella+22, Tanaka+24} and simulations \citep[e.g., ][]{Dave+20, Rey+20, Remus+25} indicate that quiescent galaxies can undergo major episodes of star formation after their quenching, causing rejuvenation. The detection of non-negligible fractions of cold gas in a large number of quiescent galaxies at both low- and high-$z$ \citep{Sargent+15, Spilker+18, Gobat+20,  Woodrum+22, Siegel+25} suggests that they can continue to accrete gas after their quenching, replenishing their gas content while remaining to be quiescent according to standard diagnostics \citep[e.g., ][]{Belli+17, Belli+21, Gobat+20, SR+20} that might fuel future rejuvenation. However, the origin and fate of cold gas in quiescent galaxies remains an open issue.

In general, galaxies can accrete gas from many sources such as the cosmic web, close companions, or the inter-galactic medium (IGM), and could potentially reignite star formation or fuel AGN activity. Direct evidence of gas accretion is rare and is found in galaxies across different redshifts and environments \citep[e.g., ][and references therein]{FoxDave17}. In the literature, gas inflows have been detected mainly in star-forming galaxies and are usually associated with ongoing SF \citep{Kacprzak+12, Rubin+12, Ho+17, Rubin+22, Weldon+23, Coleman_2024}. In contrast, the detection of neutral gas inflows in passive galaxies has been associated with AGN activity \citep{vanGorkom+89, Sato+09, Krug+10, Roy+21, Rupke+21}. Whether an inflow can trigger SF depends on the mode of accretion and the local physical conditions of the gas \citep{Tacchella+16, Tacchella+20, Zolotov+15}, which also depend on the source of the gas. Observations seem to indicate that there is no univocal origin for the inflowing material and that cosmic web, IGM, close galaxies, and recycled material from previous SF are all possible candidates \citep{Putman17}. Studying inflows in galaxies is fundamental to understand how gas accretion occurs, as it regulates the SF and the chemical enrichment of both the stellar and gaseous components, and thus the SFH, as well as the activity of the central AGN. In particular, finding inflows in quiescent galaxies is uncommon and studying such rare systems is essential for constraining models of galaxy quenching, as they challenge the simple picture in which the cessation of SF is accompanied by a complete shutdown of gas accretion.

In this work, we report the spectroscopic detection of neutral gas inflow into a long-quenched massive galaxy at $z_{\rm spec} = 2.6576 \pm 0.0003$ observed with the James Webb Space Telescope (JWST). Remarkably, this galaxy shows neither ongoing SF nor strong evidence for AGN activity, making it a unique study case for the coexistence of gas accretion and quenching. Interestingly, this galaxy resides in an overdensity, providing an opportunity to investigate how the environment influences the gas cycle in quiescent galaxies at early cosmic times. The JWST data allow us to characterize the stellar population properties of the galaxy and the properties of the inflowing material, and to investigate its origin. In Sect. \ref{sect:data} we present the data. In Sect. \ref{sect:methods} we describe the methods we used for the analysis. In Sect. \ref{sect:results} we present our results and discuss them in Sect. \ref{sect:discussion}. Finally, in Sect. \ref{sect:conclusions} we present our conclusions. Throughout this paper, we assume a \citet{chabrier} IMF and adopt a flat $\Lambda$CDM cosmology with H$_0 = 70$ km s$^{-1}$ and $\Omega_{\rm m} = 0.3$.

\section{Data}\label{sect:data}

\begin{figure*}
\centering
\includegraphics[width=\textwidth]{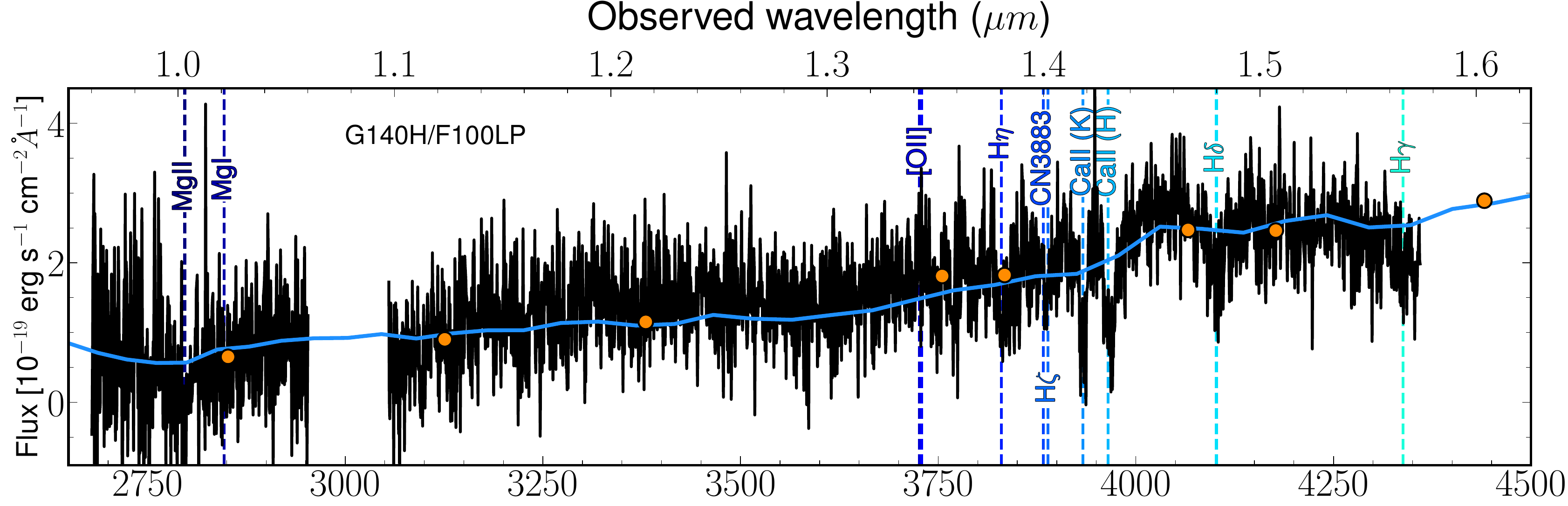}
\includegraphics[width=\textwidth]{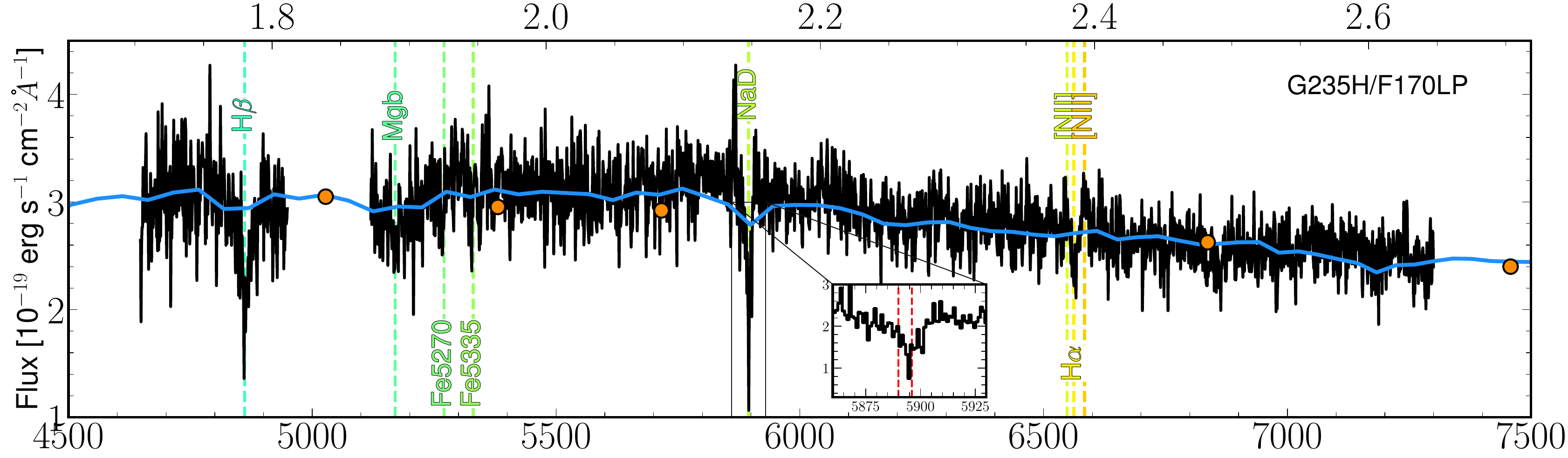}
\includegraphics[width=\textwidth]{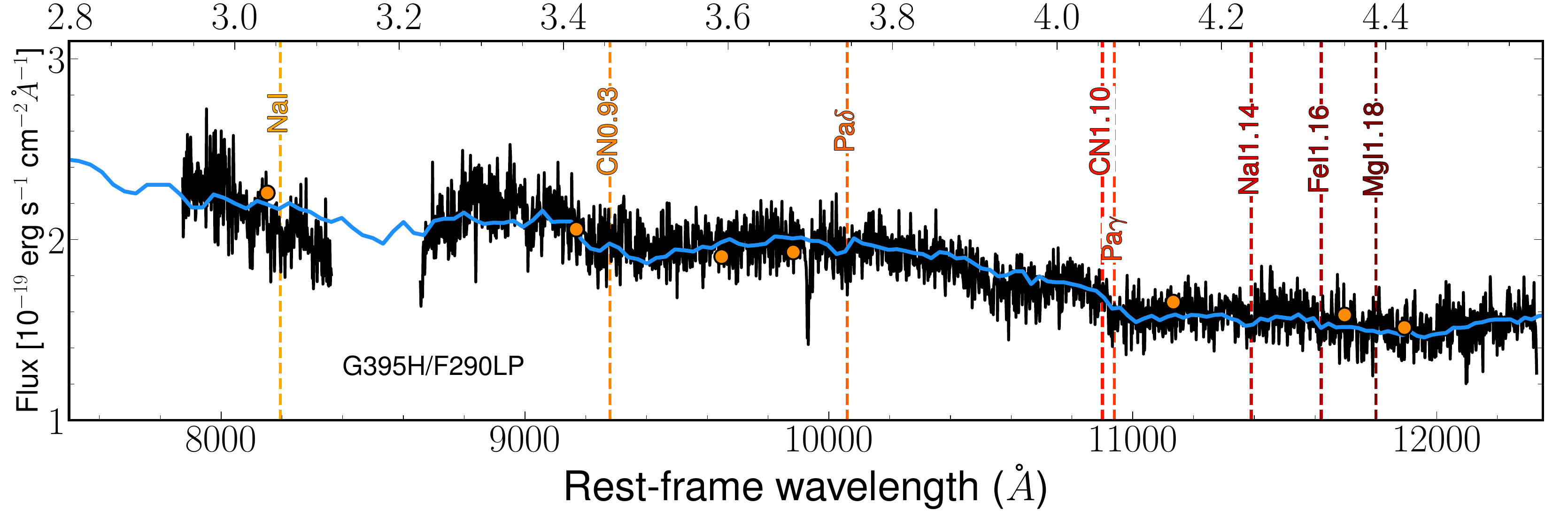}
\caption{JWST spectrum and photometry of GLASS-180009. The top, middle, and bottom panels show the wavelength regions corresponding to the G140H/F100LP, G235H/F170LP, and G395H/F290LP gratings, respectively. In each panel, the grating and PRISM spectra are plotted with black and blue curves, respectively, while the photometric data are shown as orange circles. The two spectra are rescaled to the photometry. The missing wavelength regions of the grating spectrum are the detector gaps. The main absorption and emission lines detected are highlighted with dashed vertical lines. In the middle panel, we show a zoom-in of the NaD absorption feature from the grating spectrum. The vertical red dashed lines show the expected wavelengths of the two NaI absorption lines at $\lambda \lambda \, 5890, 5896 \, \AA$, while the observed features are evidently redshifted.}\label{fig:spec}
\end{figure*}

We selected our target galaxy from the GLASS-JWST Early Release Science program \citep{glass, glass2}. The galaxy, having GLASS ID 180009 (RA: 3.605219, Dec: -30.39540; hereafter GLASS-180009), was identified as quiescent by \citet{Marchesini+23}. The galaxy was initially selected from the HFF-DeepSpace catalogs in the Abell 2744 cluster pointing constructed by \citet{Shipley+18}, having photometric redshift $> 2$, magnitude $m_{F160W}<26$, stellar mass $> 10^8 M_\odot$, and specific star formation rate $\log(\rm{sSFR}) < -9.7$. Then, it was identified as a quiecent candidate from the UVJ diagnostic and by the strong Balmer/4000 $\AA$ break. Finally, \citet{Marchesini+23} confirmed its quiescent nature by fitting the NIRCam photometry and JWST-NIRISS grism slitless spectroscopy. Follow-up JWST observations provided further photometric and spectroscopic data, as detailed below. We show the NIRSpec spectrum and the HST+NIRCam photometry of GLASS-180009 in Fig. \ref{fig:spec}. In Appendix \ref{app:calibration}, we show that the spectrum does not suffer wavelength calibration issues and describe how we rescaled it to the photometry.

We took the photometric data from the JWST UNCOVER Treasury Program \citep{Bezanson+24} and the Medium Bands Mega Science survey (MegaScience, \citealt{Suess+24}). The combined catalog, described in \citet{Weaver+24}, includes imaging for all the 20 NIRCam wide and medium bands. Specifically, UNCOVER obtained JWST/NIRCam broadband imaging of the Abell 2744 cluster in the NIRCam broad bands F115W, F150W, F200W, F277W, F356W, F410M, and F444W and MegaScience imaged the same footprint in two additional broad bands, F070W and F090W, and the medium bands F140M, F162M, F182M, F210M, F250M, F300M, F335M, F360M, F430M, F460M, and F480M. The reduction pipelines were presented in \citet{Bezanson+24} and \citet{Suess+24}. Additionally, the catalog includes HST observations in the filters F435W, F606W, F814W, F105W, F125W, F140W, and F160W. Throughout this work, we adopt the data products of the data release 4, which includes corrections for gravitational lensing effects \citep{Furtak+23} with updated spectroscopic redshifts \citep{Price+25}. The catalog ID for GLASS-180009 is 27482. 

We considered NIRSpec spectroscopic observations with the high-resolution gratings (R$\sim2700$) of JWST from GLASS, described in \citet{Mascia+24}, and with the PRISM (R$\sim100$) from UNCOVER, described in \citet{Price+25}. In this work, we used the reduced data from the DAWN JWST archive\footnote{Available from \url{https://dawn-cph.github.io/dja/}.} \citep{B24}. The median signal-to-noise ratios (S/N) of the high-resolution gratings are $\sim5, \, 18, \, \mbox{and } 20 \, \AA^{-1}$, for G140H/F100LP, G235H/F170LP, and G395G/F290LP, respectively. 

We derive the effective radius of the major axis ($R_e$) and S\'{e}rsic index ($n$) by fitting the NIRCam imaging in the F277W filter using \texttt{pysersic} \citep{pysersic}, as described in appendix \ref{app:pysersic}, for which we estimate $R_e = 1.05^{+0.01}_{-0.01}$ kpc and $n=2.15^{+0.01}_{-0.01}$. Using early data from GLASS, \citet{Marchesini+23} estimated a stellar mass of $\log(M_*/M_\odot) = 10.59^{+0.11}_{-0.05}$ (corrected for magnification effects), assuming a \citet{chabrier} stellar initial mass function (IMF). We note that, in Appendix \ref{app:test_sps}, we re-fit the stellar mass using both the spectrum and the additional photometric data and found a consistent value for the stellar mass ($4.6^{+0.1}_{-0.1} \times 10^{10} M_\odot$). 

\section{Analysis}\label{sect:methods}

\subsection{Integrated stellar velocity dispersion}\label{sect:kinematics}

We estimated the stellar velocity dispersion ($\sigma_*$) from the full spectral fitting of the high-resolution spectrum using \texttt{pPXF} \citep{ppxf_2023}. As input templates, we considered the EMILES simple stellar population (SSP) models \citep[][see the next section for a more detailed description]{EMILES} and broadened them to match the resolution of the grating spectra. We excluded from the fit wavelengths below $1.2 \, \mu m$ ($\sim 3500\, \AA$ rest frame), due to the low S/N ($< 3 \, \AA^{-1}$). We also mask Na D since it mostly originates in the neutral gas inflow, as discussed below. We then performed a first fit of the spectrum with \texttt{pPXF} and computed the standard deviation of the residuals ($\sigma_{\rm std}$). Then, we masked all the spectral pixels deviating more than $3\sigma_{\rm std}$ from the previous best fit and performed a second fit. We consider the results of this second fit as our best-fitting solutions. 

To estimate the errors, we performed a wild bootstrapping of the residuals by resampling the best-fitting spectrum with the noise Gaussianly re-distributed from the residuals. For each iteration, we performed 100 realizations and considered the standard deviation as the relative error. To account for errors due to the correlation in wavelength, in each iteration, we shuffled residuals over windows about 500 \AA \, wide. All the fits are performed including additive polynomials of degree 10 to account for possible bad subtraction of the background. We present the results of our kinematic fits in Sect. \ref{sect:dynamical}.

\subsection{Stellar population modeling}\label{sect:spmodeling}

We estimated the stellar population properties from the full spectral fitting using \texttt{pPXF} fed with the EMILES models. Specifically, we adopted the `baseFe' models computed with the BaSTI isochrones \citep{basti} and a \citet{chabrier} initial mass function (IMF). We considered models with ages $\leq2.5$ Gyr, since the age of the Universe at the redshift of the galaxy is $\simeq 2.4$ Gyr, and older than 0.07 Gyr, because the EMILES models with younger ages have lower accuracy \citep{EMILES}. In Appendix \ref{app:test_sps} we show that including models with younger ages does not change the results. For the metallicity, [M/H], we considered models with values between $-1.79 \; \mbox{and} \, +0.26$ dex.

We set up our fits as follows. We limited the fit to the rest frame wavelength region $3500-12300 \, \AA$, also excluding the NaD line, and fixed the velocity dispersion of the stars to our best-fitting $\sigma_*$ (Sect. \ref{sect:dynamical}). We simultaneously fit, as independent kinematic components, the stars and the gas emission lines, parametrized as Gaussians. In particular, we fit the [OII] doublet ($\lambda = 3726, 3728 $ \AA), HeI ($\lambda = 5877$ \AA) the [OI] doublet ($\lambda = 6302, 6366$ \AA), the [NII] doublet ($\lambda = 6718, 6733$ \AA) and the Balmer lines\footnote{We verified that other typical emission lines falling in the observed wavelength range (e.g., [NeII], [SII], etc.) are undetected or have relative errors larger than $100\%$.}. For the stellar component, we also fit a \cite{Calzetti00} reddening curve. 

We obtained the \texttt{pPXF} best-fitting model as follows. We performed a first fit of the spectrum and computed the standard deviation of the residuals ($\sigma_{\rm std}$). Then, we performed a second fit while masking the spectral pixels deviating more than $3\sigma_{\rm std}$ from the previous best fit and performed a second fit. We consider the results of this second fit as our best-fitting solutions. We calculate the mass-weighted age and metallicity as:

\begin{subequations}\label{eq:age}
\begin{align}
{\rm log}_{10}{\rm Age} = \frac{\sum_i  w_i \rm{log_{10} Age}_i}{\sum_i w_i} \\ 
{\rm [M/H]} = \frac{\sum_i w_i {\rm [M/H]}_i}{\sum_i  w_i}
\end{align}
\end{subequations}

\noindent where $w_i$ is the best-fitting weight of the $i$-th input template and the sums are performed on all the input templates. 

To estimate the errors of age, metallicity, and dust attenuation, we performed a wild bootstrapping of the residuals. To this aim, we performed 100 realizations of each spectrum and shuffled the residuals over windows $\sim500 \, \AA$ wide every time, to account for wavelength correlated errors, and Gaussianly re-distributed (i.e., added) the shuffled residuals to the best-fitting spectrum. For each realization, we estimated the stellar population parameters and consider the standard deviations of all the realizations as their errors. 

In addition to the stellar population parameters, we recover the SFH of GLASS-180009 in a non-parametric way using the same setup and methods described above, but allowing for regularization during the fit \citep{ppxf_2017}. The SFH is then calculated as the cumulative sum of the best-fitting mass weights. 

\begin{figure*}
    \centering
    \includegraphics[width=\textwidth]{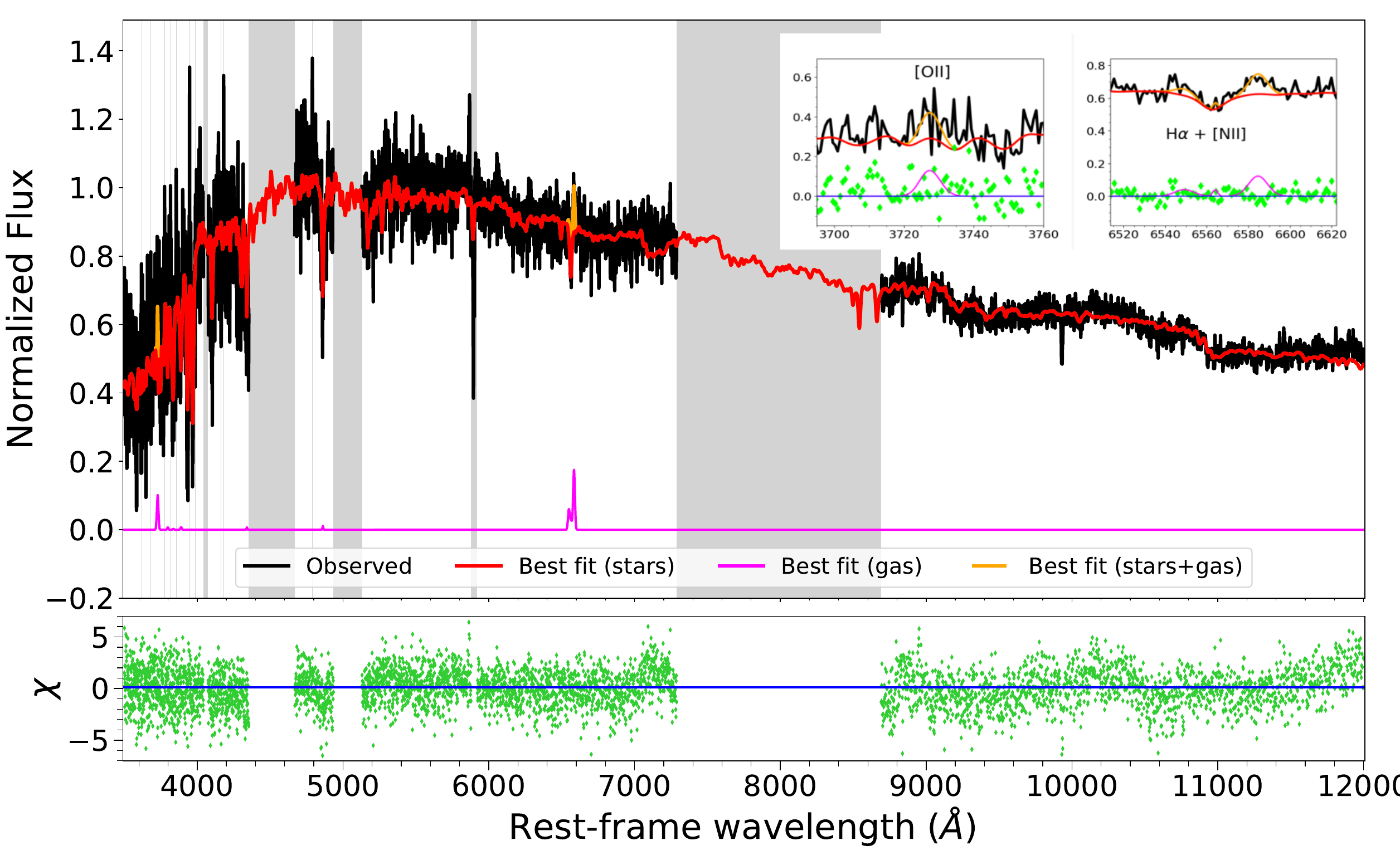}
    \caption{Best-fit of the JWST high-resolution spectrum of GLASS-180009. Top panel: the black curve is the observed spectrum; the red curve is the best-fit stellar spectrum; the magenta curve is the best fit of the gas emission lines; the orange curve is the combined best fit of the stars and gas; the gray shaded regions are the spectral regions masked during the fit. The two insets in the upper right corner show the zoom-in of the [OII] and H$\alpha$+[NII] line fits. Bottom panel: the green diamonds, $\chi$, are the difference between the observed spectrum and best-fit model spectrum divided by the errors. The blue horizontal line is the median $\chi$ value.}
    \label{fig:ppxf_bf}
\end{figure*}

\begin{figure}
    \centering
    \includegraphics[width=\columnwidth]{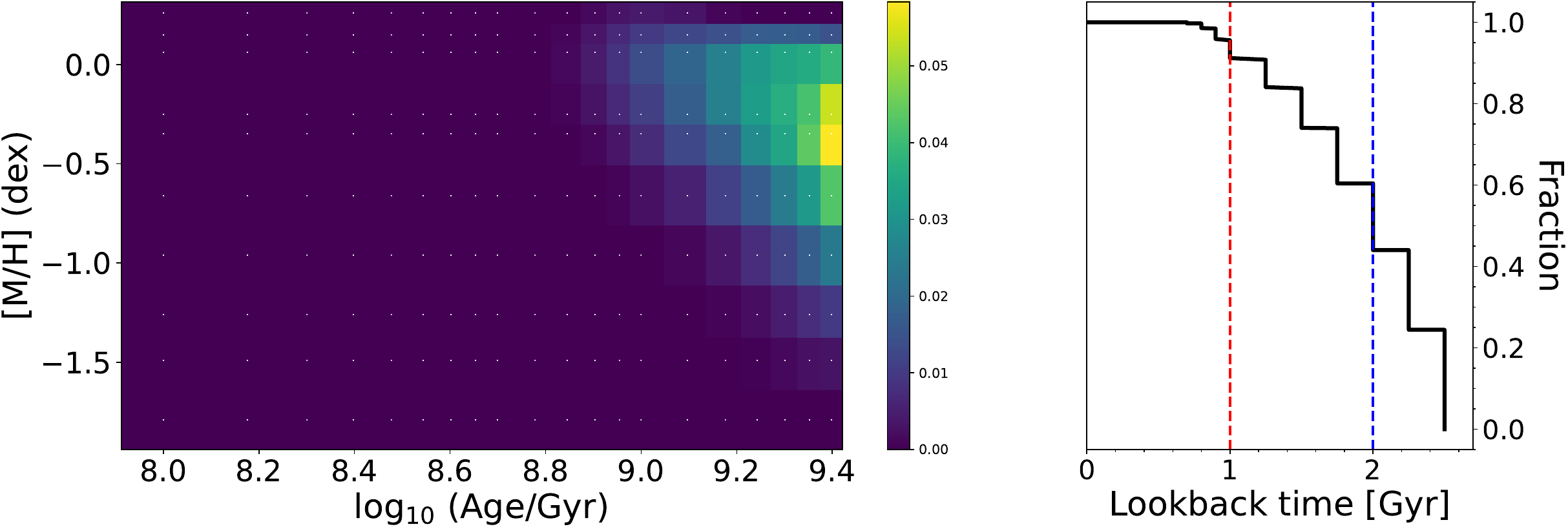}
    \caption{Left panel: map of the mass weights assigned by pPXF to each input template of given age and metallicity from the regularized fit. Right panel: fraction of mass weight as a function of the lookback time. The SFH is calculated as the cumulative sum of the weights. The red and blue lines indicate $t_{50}$ and $t_{95}$, respectively.}
    \label{fig:sfh}
\end{figure}

We show the best-fitting spectrum and the SFH in Figs. \ref{fig:ppxf_bf} and \ref{fig:sfh}, respectively. We present the results in section \ref{sect:sfh}. The best-fitting stellar population properties are reported in Table \ref{tab:ssp_params}.

\begin{deluxetable*}{c|c}
\tablewidth{0pt}
\tablecaption{Stellar population properties and emission lines estimated from the full spectral fitting with \texttt{pPXF} adopting the EMILES models.\label{tab:ssp_params}}
\tablehead{
\colhead{Parameter/Unit} & \colhead{Value}
}
\startdata
$\sigma_*$ / km s$^{-1}$  & $233\pm26$\\
Age / Gyr  & $1.74\pm0.18$\\
Metallicity / dex & $-0.26\pm0.09$ \\
$A_V$ / mag & $0.24\pm0.06$\\
$\rm{F(H\alpha)}$ / $10^{-20} \, \rm{erg} \, s^{-1} \, cm^{-2}$ & $3.2\pm2.2$ \\
$\rm{F(H\beta)}$ / $10^{-20} \, \rm{erg} \, s^{-1} \, cm^{-2}$ & $1.4\pm1.2$ \\
$\rm{F(H\gamma)}$ / $10^{-20} \, \rm{erg} \, s^{-1} \, cm^{-2}$ & $0.9\pm0.8$ \\
$\rm{F([OII])}$ / $10^{-19} \, \rm{erg} \, s^{-1} \, cm^{-2}$ & $4.1\pm1.4$ \\
$\rm{F([OI])}$ / $10^{-19} \, \rm{erg} \, s^{-1} \, cm^{-2}$ & $1.6\pm0.5$ \\
$\rm{F([NII])}$ / $10^{-19} \, \rm{erg} \, s^{-1} \, cm^{-2}$ & $6.1\pm0.8$ \\
$\rm{F(HeI)}$ / $10^{-20} \, \rm{erg} \, s^{-1} \, cm^{-2}$ & $0.0\pm0.8$ \\
\hline
SFR($H\alpha$) / $M_\odot$ yr$^{-1}$ & $0.02\pm0.01$\\
SFR($[OII]$) / $M_\odot$ yr$^{-1}$ & $0.19\pm0.07$\\
\hline
$t_{50}$ / Gyr & $2.00\pm0.25$\\
$t_{95}$ / Gyr & $1.00\pm0.15$\\
\enddata
\tablecomments{Age and metallicity are mass-weighted. Most of the other Balmer lines are detected but are very low and consistent with zero flux, so we do not report them in this table. The HeI emission is not detected; we explicitly report it here because this emission line would affect the NaD measurement, if detected.  The SFR values are derived from the emission lines (Sect. \ref{sect:sfh}). The values of $t_{50}$ and $t_{95}$ are derived from the SFH, reconstructed from the regularized fits.}
\end{deluxetable*}

\subsection{Modeling the NaD absorption lines tracing the neutral gas inflow}\label{sect:voigt}

Inspection of the NaI doublet at $\lambda 5890, 5896 \AA$ (Fig. \ref{fig:spec}) shows clear redshifted absorption which is caused by neutral gas in inflow. We then derived the properties of the inflowing gas by modeling the NaD absorption lines with \texttt{VoigtFit} \citep{voigtfit}. The code models the lines with a Voigt profile and fit the velocity, $v$, the broadening due to thermal and turbulent motions, $b$, and the column density of the absorber, $N_{NaI}$. Before fitting the observed spectrum, we corrected it for stellar absorption by subtracting the \texttt{pPXF} best-fitting stellar population model spectrum. We note that we do not detect any HeI emission (see Table \ref{tab:ssp_params}), which in principle could affect the Na absorptions.

\begin{figure}
    \centering
    \includegraphics[width=\columnwidth]{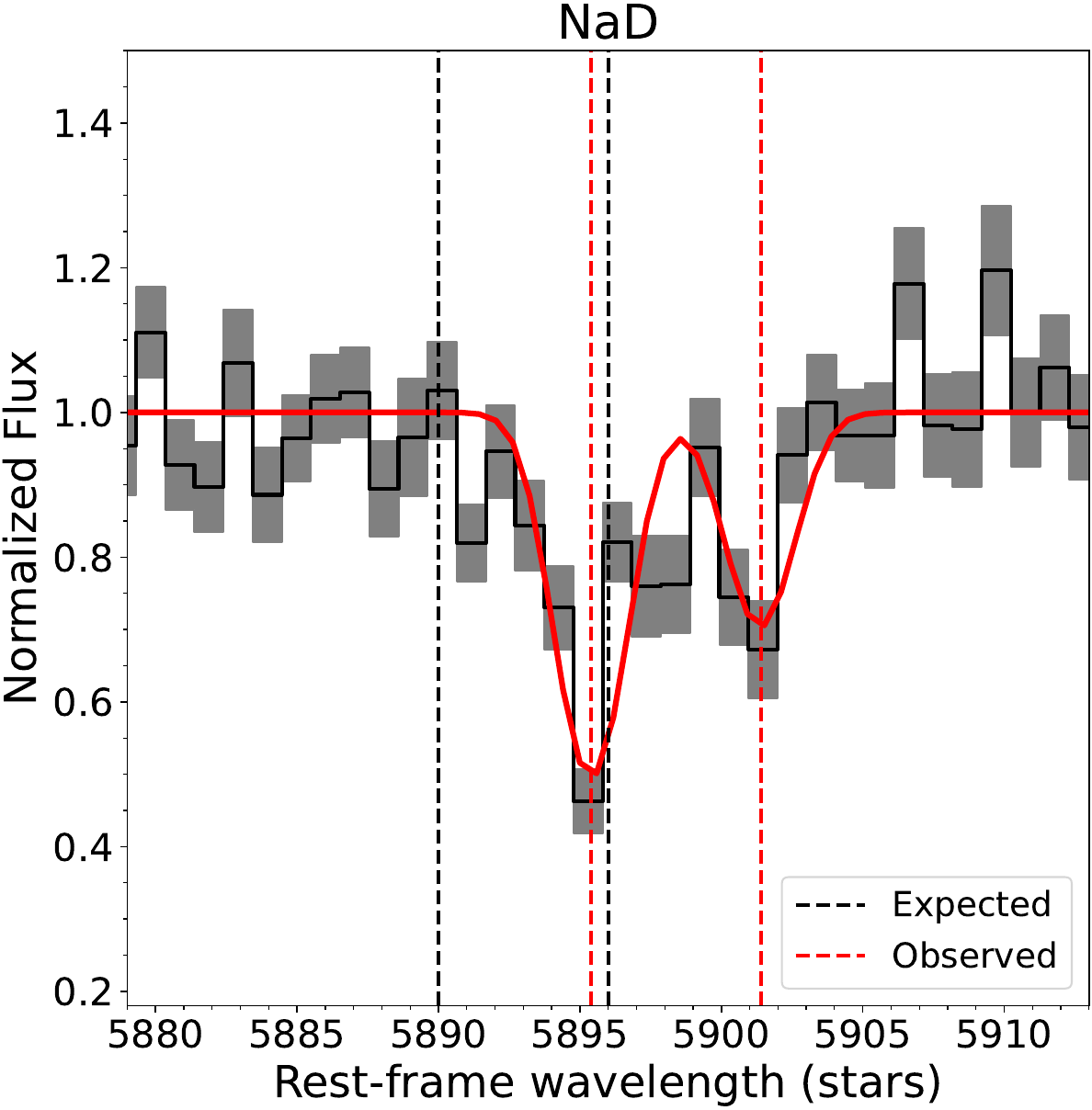}
    \caption{NaD absorption features of the inflowing neutral gas. The black curve is the observed spectrum corrected for the stellar continuum. The gray regions indicate the errors on the flux. The red curve is the best-fitting solution of \texttt{VoigtFit}. The black (red) vertical lines indicate the wavelengths of the two expected (observed) NaI absorption lines.}
    \label{fig:nad}
\end{figure}

To assess the best-fitting parameters and errors, we used a Bayesian Markov Chain Monte Carlo (MCMC) approach. We implemented MCMC sampling using the ensemble sampler implemented in \texttt{emcee} \citep{emcee} to estimate the posterior probability distribution of the model parameters. We initialized the sampler with 50 walkers and ran the MCMC chains for 5000 steps, discarding the first 1500 steps as burn-in. The likelihood function was defined as $\ln \mathcal{L} = - \frac{1}{2} \chi^2$, where $\chi^2$ is returned by \texttt{VoigtFit} for each given set of input parameters, $v$, $b$, and $\log(N_{NaI})$. We adopted flat priors for all parameters. We consider the best-fitting parameters of the marginalized posterior distributions as the median values, and the 16th and 84th percentiles as the errors. We show the best fit of the NaD in Fig. \ref{fig:nad} and report the best-fitting parameters in Table \ref{tab:inflow}. We present our results in more detail in Sect. \ref{sect:inflow}.

\begin{deluxetable*}{c|c}\label{tab:inflow}
\tablewidth{0pt}
\tablecaption{Properties of the inflowing neutral gas estimated from the fit of the NaI doublet at $\lambda \,5890, 5896 \, \AA$.}
\tablehead{
\colhead{Parameter / Unit} & \colhead{Value}
}
\startdata
$v$ / km s$^{-1}$  & $278^{+79}_{-79}$\\
$b$ / km s$^{-1}$  & $78^{+26}_{-25}$\\
$\log(N_{NaI})$ / cm$^{-2}$  & $13.02^{+0.03}_{-0.03}$\\
\hline
$\log(N_{H})$ / cm$^{-2}$  & $20.66^{+0.03}_{-0.03}$\\
$M_{in} \, / \, M_\odot $ & $ 1.6^{+0.1}_{-0.1} \times 10^8$\\
$\dot{M}_{in} \, /  M_\odot$ yr$^{-1}$ & $19_{-7}^{+6}$ \\
\enddata
\tablecomments{Errors take into account uncertainties in the redshift and the spectral sampling. $\log(N_{H})$ is calculated using eq. \ref{eq:nh}. $M_{in}$ and $\dot{M}_{in}$ are calculated using eqs. \ref{eq:min}.}
\end{deluxetable*}

\section{Results}\label{sect:results}

\subsection{Dynamical properties}\label{sect:dynamical}

From the kinematic fits, we find best-fitting $\sigma_* = 233\pm26$ km s$^{-1}$. To have an independent check and to account for possible wavelength calibration issues among the different NIRSpec gratings \citep[e.g., ][]{DR3}, we re-estimated $\sigma_*$ by fitting only the G140H/F100LP spectrum with the MILES stars \citep{MILES} using the same methods described in Sect. \ref{sect:kinematics} and found a consistent best-fitting $\sigma_* = 227\pm18$ km s$^{-1}$. Finally, we note that we find a consistent velocity dispersion when fitting the spectrum with \texttt{Bagpipies} (see Appendix \ref{app:test_sps}). We adopt the $\sigma_*$ derived from the fit of the entire spectrum with \texttt{pPXF} fed with the EMILES models as our reference value. 

From the stellar velocity dispersion, we infer the dynamical mass of GLASS-180009  $M_{\rm{dyn}} = k(n)  R_e \, \sigma_e^2/ G \simeq 9.9 \times 10^{10} \, M_\odot$, assuming $k(n) \approx 8.87 - 0.83 \, n + 0.024 \, n^2 $ \citep{Cappellari+06} and $\sigma_e \approx \sigma_*$\footnote{This is justified because the effective radius of GLASS-180009, $R_e \simeq 0.13''$, is comparable (lower than) the (circularized) aperture of the microshutter $R_{\rm {ap}}\simeq 0.17''$, so we do not apply any aperture correction \citep{Jorgensen+95, Cappellari+06}.}. This implies a dark matter fraction $M_{\rm{DM}}/M_{\rm{dyn}} = 1-M_*/M_{\rm{dyn}}\simeq 0.61$, namely a factor of $\sim1.5$ larger than the stellar mass, which is unusually large at $z>2$ \citep[e.g., ][]{vandesande+13, Beifiori+14, Tortora+14, lovell+18, Mendel+20}. Alternatively, under the assumption of the galaxy being baryon-dominated within the effective radius, the comparison between the dynamical and the stellar masses suggests that the IMF is more bottom-heavy than the \citet{chabrier} IMF. Assuming a variation of the IMF with velocity dispersion \citep[e.g., ][]{cappellari_imf, Conroy+13, LaBarbera+13}, we would need a Salpeter-like IMF to have $ M_* \approx M_{\rm{dyn}}$. This is in agreement with recent results by \citet{Slob+25} based on a sample of 15 massive quiescent galaxies at $z\sim2$. However, we note that the S\'{e}rsic index of GLASS-180009 is relatively low ($n = 2.15$) compared to the typical values of local quiescent galaxies with comparable $\sigma_*$ and mass \citep[e.g., ][]{Zahid+17}. This might indicate the presence of stellar rotation \citep[as also found in ][]{Slob+25} that would account for part of the broadening. Then $\sigma_*$ and $M_{\rm{dyn}}$ might be overestimated. Furthermore, we note that a variation of the IMF with the radius can have a significant impact on the estimate of $M_*$ and $M_{\rm dyn}$ \citep{MN+15, vDC+17, FLB+19, Bernardi+23}. Finally, we note that the relation between the dynamical mass and velocity dispersion we adopted is calibrated on local galaxies and might be different at high-$z$.

\subsection{Stellar population properties and SFH}\label{sect:sfh}

From the fit of the stellar population properties, we find that GLASS-180009 has an age of $1.74\pm0.18$ Gyr, a subsolar metallicity, [M/H] = $-0.26\pm0.09$ dex, and dust attenuation $A_V = 0.24\pm0.06$ mag. The recovered SFH suggests that the galaxy formed 50 and 95 percent of its mass at lookback times $t_{50} = 2.00\pm0.25$ Gyr and $t_{95} = 1.00\pm0.15$ Gyr, corresponding to $z_{50} = 11^{+18}_{-3}$ and $z_{95} \simeq 4.3^{+0.3}_{-0.4}$, respectively, and then quenched without any significant recent SF. We note that, according to its SFH, GLASS-180009 is an `ancient' galaxy, having formed $\geq50 \%$ of its mass at $z_{50} \geq 11$, similar to other galaxies recently discovered with JWST and possibly challenging the $\Lambda$CDM \citep{Glazebrook+24, Carnall+24, degraaff_z5, McConachie+25}, although in this case the mass at $z_{50}$ is not as high. 

In Appendix \ref{app:test_sps} we test the reliability of the derived stellar population properties and SFH. In particular, we fit the spectrum using the \citet{BC03} models (updated to version 2016; hereafter, BC16) to consider the effect of including models with very young ages (down to 0.1 Myr). In addition, we test the reliability of our fitting methods by comparing with the results from \texttt{Bagpipes} \citep{bagpipes}. In summary, we find that the BC16 models favor solutions with younger ages and higher metallicities than the EMILES models. However, this difference is not due to the inclusion of young models, as they do not contribute to the best-fitting model spectrum, but rather to the wavelength region fitted and to differences in the model libraries. Crucially, we find that the quenching time, $t_{95}$, is the same. We find consistent results when fitting the spectrum with \texttt{Bagpipes}. 

We derive the current star formation rate (SFR) from the H$\alpha$ and [OII] emission lines as follows. First, we correct the measured fluxes for dust attenuation as $F_{corr} = F \times e^{\tau_{\scriptsize{\lambda}}}$, with $\tau_\lambda = A_\lambda/1.086$ and $A_\lambda = k_\lambda A_V/R_V$, where $k_\lambda$ is the reddening curve for the gas, for which we assume a \citet{Calzetti00} law with $R_V = 4.05$, and $A_V = 0.24$ mag, as estimated from the fit. Then we calculate the luminosities of the two lines as $L = 4 \pi d_L^2 \times F_{corr}$, where $d_L = 2.2 \times10^4$ Mpc is the luminosity distance and find $L_{H\alpha} \simeq 2.3 \times 10^{39} \; \rm{erg \, s}^{-1} $ and $L_{\rm{[OII]}} \simeq 2.9 \times 10^{40} \; \rm{erg \, s}^{-1} $. Thus, we estimate the SFR from H$\alpha$ using equation 2 of \citet{Kennicutt98}, SFR$(H\alpha)$ = $7.9\times 10^{-42}\,L_{H\alpha} = 0.02\pm0.01$ M$_\odot$ yr$^{-1}$, and from [OII] using equation 4 of \citet{Kewley+04}, SFR$([OII]) = 6.58\pm1.65\times 10^{-42}\,L_{[OII]} = 0.19\pm0.07$ M$_\odot$ yr$^{-1}$, assuming a \citet{chabrier} IMF. 

We note that these values are upper limits to the actual SFR. In particular, the [OII] emission could also be due to the presence of evolved stars, such as post-AGB, or shocks \citep[e.g.,][]{Dopita+95, CidFernandes+11}. This could be probed, for instance, with the [OIII] line at $\lambda = 5007 \, \AA$ which could be used to derive the ionization level of the nebular emission and discriminate between low and high excitation. However, [OIII] is not measurable from the high-resolution spectrum because it falls within the detector gaps. From the PRISM spectrum, which covers the wavelength of the [OIII] line, we do not detect any emission either, and neither was detected from the NIRISS slitless spectrum by \citet{Marchesini+23}. Assuming that the non-detection in the PRISM is due to its low resolution spreading out the flux over the FWHM, we can estimate an upper limit to the flux of [OIII].\footnote{The minimum flux integrated over the resolution element, $\Delta \lambda = \lambda/R$, required to have a $1\sigma$ detection can be calculated as $F_{\rm min} = f_\lambda \cdot \Delta \lambda = \sigma_\lambda \cdot \Delta \lambda$, where $\sigma_\lambda = \sigma_{\rm{pix}} \cdot \sqrt{N_{\rm pix}} = \frac{f_{\rm c}}{S/N}  \cdot \sqrt{\frac{\Delta \lambda}{\delta \lambda}}$ is the noise within the resolution element, with $f_c$ and $\delta \lambda$ being the flux of the continuum and the spectral sampling, respectively.} We estimate $F_{\rm min}([OIII]) = 5.3 \times 10^{-20} \, \rm{erg} \, s^{-1} \, cm^{-2}$, above which we would detect the emission. This implies an upper limit to $\log([OIII]/H\beta) \simeq 0.6$ dex, which, combined with the relatively high ratios $\log([NII]/H\alpha) \simeq 1.3$ dex and $\log([OI]/H\alpha) \simeq 0.7$ dex, would place this galaxy in the LI(N)ER region of the BPT diagram \citep{BPT, Kewley+06, Belfiore+16}.

Given the stellar mass and SFR of GLASS-180009, we derive an upper limit to the sSFR as SFR([OII])/M$_* = 0.5\pm0.1 \times 10^{-11}$ yr$^{-1}$, which is more than 1 dex lower than the limit defined by \citet{Gallazzi+14} for local galaxies, sSFR $ < 0.2 / t_{\rm U} (z) \simeq 8 \times 10^{-11}$, where $t_{\rm U} = 2.4$ Gyr is the age of the Universe at the galaxy's redshift. This confirms the quiescent nature of GLASS-180009. We also mention that we measure a (dust-corrected) value for the spectral index D$_n$4000 \citep{Balogh+99} of $1.47\pm0.02$, which is typical of post-starburst galaxies that do not have significant star formation \citep{Kauffmann+03, Wild+09, French+18}.

Finally, we mention that the quality of the spectra is not sufficiently high to robustly constrain the elemental abundances of GLASS-180009. Furthermore, because of the relatively young age of this galaxy, the effect of elemental abundances is intrinsically hard to measure. As described in Appendix \ref{app:mgfe}, we attempted to constrain the abundance of [Mg/Fe] for which we estimate $\rm{[Mg/Fe]} = 0.11^{+0.15}_{-0.13}$ dex, suggesting possible $\alpha$-enhancement.

\subsection{Properties of the inflowing gas}\label{sect:inflow}

From the fit of the NaI doublet at $\lambda \lambda \, 5890, 5896 \, \AA$ with \texttt{VoigtFit} (Sect. \ref{sect:voigt}), we estimate\footnote{The reported errors were propagated to account for uncertainties on redshift and the spectral sampling.} $v = 278^{+79}_{-79}$ km s$^{-1}$, $b = 78^{+26}_{-25}$ km s$^{-1}$, and $\log(N_{NaI}/\rm{cm^2}) = 13.02^{+0.03}_{-0.03}$.  In Fig. \ref{fig:nad} we show the \texttt{VoigtFit} best-fitting profile.

Since the best-fitting model does not account for possible Na-enhancement or the presence of neutral gas in the galaxy, we tested these effects by manually deepening the NaD stellar features in the \texttt{pPXF} best-fitting spectrum. To do this, we used the Na-enhanced models by \citet[][hereafter CvD18]{CvD18} with [Na/H] = +0.3 dex, interpolated at the same age and metallicity as our best-fitting values. We then subtracted the Na-enhanced spectrum from the observed continuum and ran \texttt{VoigtFit} as described in Sect. \ref{sect:voigt}. As a result, we find a lower value for the column density of NaI, $\log(N_{NaI}/\rm{cm^2}) = 12.94^{+0.04}_{-0.04}$, while $v$ and $b$ are consistent. However, this small difference does not qualitatively change our results. 

We note that the rest frame JWST high-resolution spectrum also covers the MgII doublet at $\lambda \lambda \, 2796, 2803 \, \AA$ and the CaII-K line doublet at $3934 \, \AA$, two other tracers of the neutral gas \citep[e.g.,][]{Wu25, Liboni+25}. In the former case, a shift towards redder wavelengths is recognizable, but we could not perform a robust fit because of the low S/N ($\sim 2 \, \AA^{-1}$). In the latter case, the shift is not easily distinguishable because the gas absorption falls within the stellar feature. We mention that, by correcting for stellar absorption, we indeed found residuals consistent with the presence of an inflow in CaII-K. However, the stellar CaII-K absorption is sensitive to Ca abundance and is affected by H$\epsilon$, and we cannot safely constrain them with current data, so we did not attempt to derive the properties of the inflowing gas from the CaII-K line.

Following \citet{Rupke+05}, \citet{Veilleux+20}, \citet{Wu25}, and \citet{Valentino+25}, we converted $\log(N_{NaI})$ estimated from NaD into hydrogen column density using the following equation:

\begin{equation}\label{eq:nh}
    \log(N_H) = \log(N_{NaI}) - \log(X_{Na}^0)  - A_{Na} - B_{Na} 
\end{equation}

\noindent where $X_{Na}^0 \equiv N_{NaI}/N_{Na}$ is the fraction of Na that is neutral, $A_{Na}$ is the abundance of Na  relative to H in the gas phase, and $B_{Na}$ is the level of depletion of Na into dust. As described in \citet{Veilleux+20} and following \citet{Wu25} and \citet{Valentino+25}, we adopt a neutral gas fraction $X_{Na}^0 = 0.1$ \citep{Stokes78}, a Milky-Way-like sodium abundance, $A_{Na} = -5.69$, and depletion $B_{Na} = -0.95$ \citep{Savage+96}. Thus, we estimate $\log(N_H/cm^2) = 20.66^{+0.03}_{-0.03}$.

Then, following \citet{Rupke+05}, \citet{Davies+24}, \citet{Wu25}, and \citet{Valentino+25}, we estimate the inflowing mass and rate, assuming a simple spherical thin-shell model, as:

\begin{subequations}\label{eq:min}
\begin{align}
M_{in} = 1.4 \, m_p \, \Omega \, N(H)\, R^2_{in} \\
\dot{M}_{in} = 1.4 \, m_p \, \Omega \, N(H) \, R_{in} \, v_{in}
\end{align}
\end{subequations}

\noindent where $m_p$ is the mass of the proton, $\Omega$ is the solid angle subtended by the inflow, $R_{in}$ is the shell radius, and $v_{in}$ is the velocity of the inflow from the best-fit of \texttt{VoigtFit}. Following the above-mentioned papers, we set $\Omega = 0.45 \times 4\pi$, which assumes a randomly oriented inflow \citep{Wu25}, and $R_{in} = 2\times R_e$, which assumes a covering fraction for NaI of $\approx 1$ and that the physical extent of the inflowing gas is as extended as (half of) the stellar light. We estimate an inflowing mass $M_{in} = 1.6^{+0.1}_{-0.1} \times 10^8 M_\odot$ and rate $\dot{M}_{in} = 19_{-7}^{+6}$ $M_\odot$ yr$^{-1}$. 

We caution that these assumptions represent the main source of uncertainty in our analysis. For instance, studies based on outflows show that the covering fraction of NaI is typically lower than 1 and generally varies in the range $\sim 0.3 - 1$, depending on the source of the outflow \citep{Rupke+05, Martin+09, Rubin+14, Davies+24}. Furthermore, NaD is expected to be clumpy \citep[e.g., ][]{Roy+21}. Therefore, our assumptions about the properties and geometry of the neutral gas could lead to an overestimate of the inflowing mass and rate. On the other hand, both observational (e.g., \citealt{Rupke+05, Martin+09}) and theoretical studies (e.g., \citealt{Murray+07, Richter+11}) show that NaI traces only the coldest and densest gas, resulting in systematically lower inferred hydrogen column densities compared to estimates based on other lines like MgII and FeII, which trace more diffuse or ionized phases. For example, \citealt{Valentino+25} estimated values of $\log(N_H) \sim 1$ dex higher when estimated from MgII and FeII compared to NaI. This implies that we might be underestimating $\log(N_H)$, and thus $M_{in}$ and $\dot{M}_{in}$. Finally, the assumptions adopted in eq. \ref{eq:nh} also affect the estimates of $N_H$, although \citet{Moretti+25} recently confirmed the validity of these assumptions for one massive quiescent galaxy at $z\sim2$. In summary, our estimates have intrinsic uncertainties that can only be reduced with targeted observations of the neutral gas to robustly constrain the hydrogen column density and the geometry of the inflow. However, they allow us to directly compare our results with other works in the literature. 

\section{Discussion}\label{sect:discussion}

In the previous sections, we have shown that GLASS-180009 present an evident redshift of the NaD lines, indicating the presence of neutral gas inflow for which we derived the mass and rate. Despite the gas inflow, the galaxy remains quiescent, with no evidence of any significant SF burst during the last $\sim 1$ Gyr, according to its SFH. In this section, we discuss the possible sources of the inflow and the possible reasons why the inflow does not cause an ignition of the SF.

\subsection{The source of the inflowing neutral gas}

\begin{figure}
    \centering
    \includegraphics[width=\columnwidth]{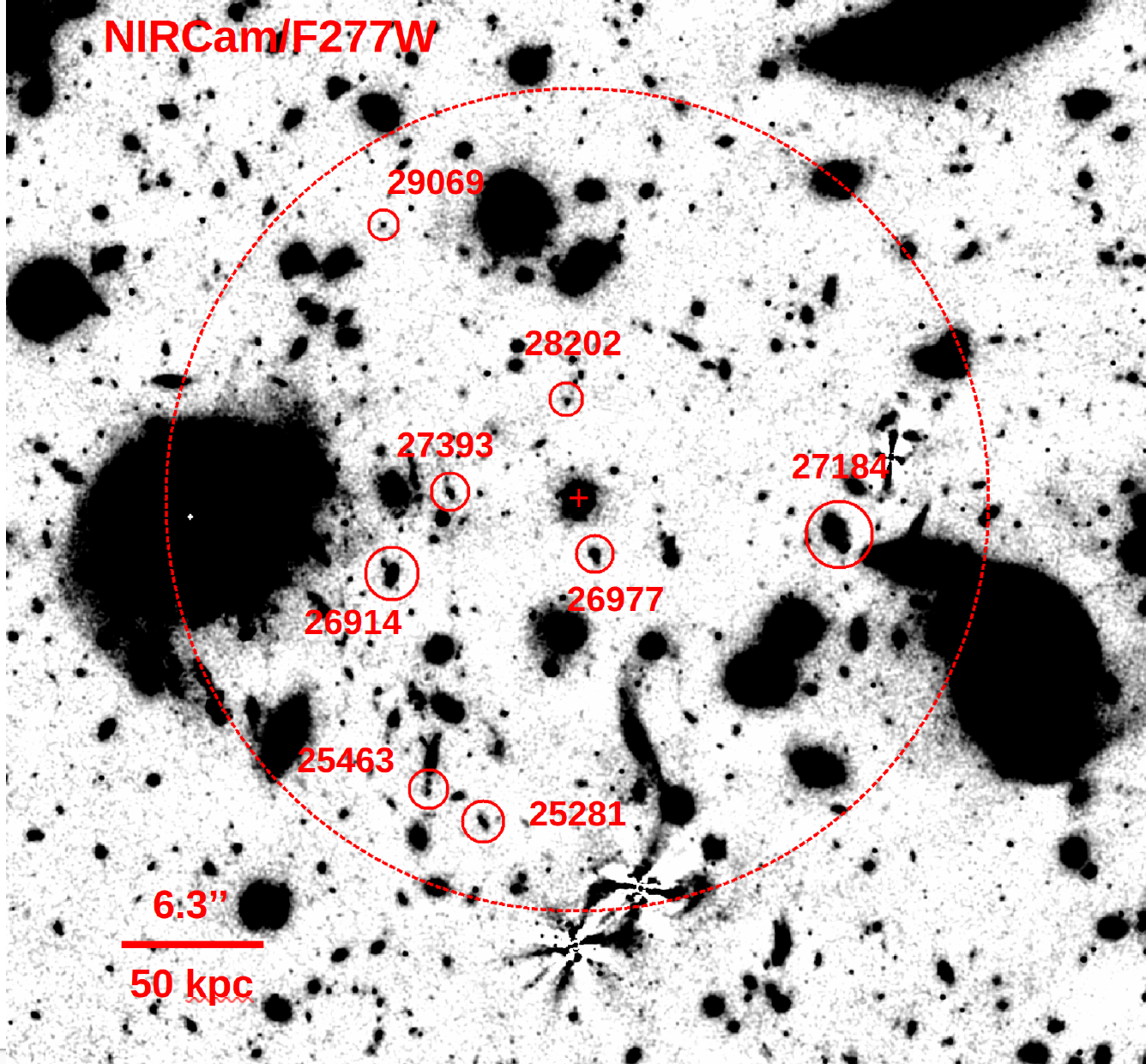}
    \caption{NIRCam imaging of GLASS-180009 in the F277W filter. The scale is shown in the lower left corner. The target galaxy is marked with a cross. The big dashed circle demarcate a surrounding region of about 150 kpc at the galaxy's redshift. The small circles highlight the 8 galaxies within 150 kpc projected distance having redshift comparable to that of our target galaxy. The UNCOVER catalog ID of each galaxy is reported close to each circle.}\label{fig:environment}
\end{figure}

Different sources of neutral gas can be responsible for the observed inflow, such as satellite galaxies through tidal streams or mergers, the IGM or cosmic filaments, and galactic fountains \citep{FoxDave17}. Since GLASS-180009 is residing in an overdensity \citep{Watson+25, Pan+25}, satellite galaxies represent viable sources for the neutral gas. To test this hypothesis, we checked the environment of GLASS-180009 in the JWST imaging. In Fig. \ref{fig:environment} we show a $50'' \times 50''$ (about $400 \times 400$ kpc) zoom-in image of the NIRCam/F277W filter from UNCOVER centered at our target galaxy. We looked for close companions of GLASS-180009 in the UNCOVER/MegaScience catalog by selecting galaxies with photometric redshifts between 2.6 and 2.8 (i.e., approximately $\pm 0.1$ from our target) and with projected distances within 150 kpc. We found 10 galaxies surrounding GLASS-18009. We compared the photometric redshifts of these galaxies with those in the NIRISS catalog \citep{Watson+25} and confirmed the proximity only for two galaxies (GLASS IDs 25281 and 26914), while two other galaxies have significantly different redshifts, so we excluded them; the remaining 6 galaxies do not have redshift measurements in the NIRISS catalog. We show the 8 surrounding galaxies in Fig. \ref{fig:environment}. From the DR4 catalog of the photometric SED fitting \citep{Wang+24}, we found that all these galaxies have low stellar masses, $\log_{10}(M_*/M_\odot) = 6.8 - 9.2$, young ages, $0.1-1$ Gyr, and relatively high sSFR, $\log_{10}(\rm{sSFR/Gyr}^{-1}) = 1-10$. Among these, only two galaxies, 26977 and 26914, have a stellar mass higher than the mass of the inflowing gas, $10^{8.2} M_\odot$. In particular, 26977 appears as the closest surrounding galaxy, having projected distance $\lesssim 20$ kpc. If the neutral gas inflow is a consequence of an ongoing wet minor merger, then this galaxy represents the best source candidate. If instead it is due to tidal streams, it is likely that more than one galaxy is contributing to the total mass of the inflowing gas. We note that \citet{Davies+24} found a passive galaxy with evidence of neutral gas inflow, highlighting the presence of a close companion that is possibly fueling the inflow, similar to our case.

To further investigate the environment, we considered the ALMA observations of the Abell2744 cluster at 244-274 GHz recently released by \citet{dualz}. We show the ALMA observations of the same sky region shown in Fig. \ref{fig:environment} in Appendix \ref{app:alma} (Fig. \ref{fig:alma-dualz}). Unfortunately, these observations are dominated by the noise and do not show any clear detection of gas around GLASS-180009 or the surrounding galaxies. Therefore, targeted observations for neutral gas are needed to further address the origin of the inflow. This would also allow us to probe the IGM or cosmic filaments as other sources of the inflow, which cannot be ruled out on the basis of current data. 

Finally, we consider the possibility that the inflow originates from a galactic fountain, i.e. from the recycled material of the SF. There are two main arguments against the galactic fountain. First, they are usually associated with significant SF activities. Both observations (e.g., \citealt{Fox+19}) and models (e.g., \citealt{Marasco+12}) indicate that the rate of neutral gas inflow due to galactic fountains is of the same order of the SFR. For example, \citet{Davies+24} recently reported two star-forming galaxies with evidence of neutral gas inflows (having masses and rates very similar to those of our target galaxy), potentially indicating galactic fountains \citep[see also, e.g.,][]{Rubin+12, Weldon+23}. However, the SFR ($\leq 0.2 \, M_\odot \, \rm{yr}^{-1}$) of GLASS-180009 is at least $2$ dex lower than the mass accretion rate of the gas ($\sim20 \, M_\odot \, \rm{yr}^{-1}$). Even assuming a delayed re-accretion of the ejected material, the typical galactic fountain return time scales are of the order of a few hundreds of Myr \citep{Oppenheimer+08, Spitoni+09, Spitoni+13, Fox+19, Grand+19}, while GLASS-180009 stopped forming stars $\sim 1$ Gyr before observations, according to its SFH. Secondly, observations indicate that galactic fountains typically exhibit hydrogen column densities of $\log(N_{\mathrm{H\,I}}/\mathrm{cm}^{-2}) \sim 18\text{–}20$ and sodium column densities of $\log(N_{\mathrm{Na\,I}}/\mathrm{cm}^{-2}) \sim 12$  \citep[e.g.,][]{Bekhti+08, Putman+12, Rubin+22}, i.e., at least 1 dex lower than the densities we estimated. Therefore, it is unlikely that the neutral gas inflow we detect in GLASS-180009 is due to a galactic fountain.

\subsection{The fate of the inflowing gas}

An inflow of neutral gas in galaxies is expected to trigger and fuel SF. However, the ignition of SF is sensitively dependent on the local conditions of the gas. There are several possible reasons why the inflow we detect in GLASS-180009 does not trigger the SF. First, the mass of the gas might be too low to ignite significant SF. According to the empirical relation defined by \citet{Gallazzi+14} for local galaxies, a galaxy is non-passive if its sSFR$\geq 0.2/ t_U \simeq 8 \times 10^{-11} \, \rm{yr}^{-1}$ at the galaxy's redshift. For a stellar mass of $\simeq 4 \times 10^{10} M_\odot$, as GLASS-180009, this implies $\rm{SFR}_{min}\geq 3 \, M_\odot \, \rm{yr}^{-1}$. Using the scaling relations derived by \citet{Tacconi+18}, we can estimate the expected depletion time of the gas in a galaxy at $z\sim 2.7$ with the stellar mass of GLASS-180009, namely $t_{dep}\simeq0.5$ Gyr. Then, we can estimate the minimum mass of the gas required to have $\rm{SFR}_{min}$ as $M_{\rm{gas, min}} = \rm{SFR}_{\rm{min}} \times t_{depl} \simeq 2 \times 10^9 \, M_\odot$, which is 1 dex larger than the mass of the inflowing neutral gas we estimate. Therefore, the reason why GLASS-180009 remains passive could be that the mass of the gas is too low to enhance significant SF. However, this assumes that the total gas mass fueling the SF is that of the inflow, which might not be the case if some gas is also residing in the galaxy.

Another reason why we do not detect a significant SFR might be that the gas density is too low to efficiently form stars. Several works have shown that in nearby galaxies there is a quite sharp transition at $\log(N_H) \sim 21$, below which the hydrogen is predominantly neutral, while the molecular $H_2$ is dominant at higher column densities \citep{Blitz+06, Leroy+08, Bigiel+08, Krumholz+09}. These works consistently show that the transition in $\log(N_H)$ corresponds to a gas surface density $\Sigma_{gas} \sim 10 \, M_\odot \, \rm{pc}^{-2}$, which is exactly the critical threshold found empirically by \citet{Kennicutt98} above which star formation occurs because the gas is gravitationally unstable, according to the theoretical Toomre criterion (see also \citealt{Kennicut89, Martin+01}). For GLASS-180009, we estimate $\log(N_H/\rm{cm}^2) \simeq 20.7$, suggesting that the density could be too low to efficiently convert gas into stars. However, considering the uncertainties on $\log(N_H)$ discussed in Sect. \ref{sect:inflow}, observations of the HI and H$_2$ are needed to further probe this argument.

On the other hand, we point out that the works mentioned above discussing the density thresholds are based on galaxies whose gas is already settled. For an inflowing gas, instead, we might expect it to compress and ignite the SF as it falls into the galaxy due to compressive turbulence, shocks, or converging flows, on characteristic time scales of the order of $10-100$ Myr, depending on the conditions of the gas \citep{Hollenbach+71, Jura75, Bergin+04, Glover+07}. Assuming a constant inflow mass and rate of neutral gas for GLASS-180009, we can estimate the overall time scale for the inflowing gas as $t_{in} \equiv M_{in}/\dot{M}_{in}\simeq 16$ Myr. This is consistent but relatively short compared to the time scales at which the infalling gas is expected to compress the gas and form stars. Therefore, the SFR could be low because the time scale of the infall is too short to compress the gas and ignite the SF.

We note that if the inflow is fueling some low-level SF this is clear direct evidence that galaxies can accrete cold gas and keep forming stars long after their quenching, while remaining quiescent according to the standard diagnostics \citep[e.g., ][]{Belli+17, Belli+21, Gobat+20, SR+20, Zhu+25}. If the gas accretion and star formation are sustained for several Gyr through continuous or episodic events, the younger populations formed would account for a non-negligible fraction of the total stellar mass we measure at lower redshifts, as inferred from their SFHs \citep[e.g.,][]{Tacchella+22, Bevacqua+24}. Alternatively, if no SF is ignited at all, the accreted gas might end up replenishing the neutral gas content of the galaxy, in line with the large dust content found in high-redshift quiescent galaxies \citep[e.g., ][]{Siegel+25, Lorenzon+25}

Finally, it is possible that the neutral gas is fueling the AGN, assuming a central accretion of the gas. In fact, a link of the neutral gas inflow with the AGN activity in passive galaxies has already been reported in the literature \citep{Sato+09, Roy+21}. However, the AGN emissions reported in these works are much stronger and the accretion rates are much lower ($\sim 0.1-5 \, M_\odot \, \rm{yr}^{-1}$). Even assuming a very low efficiency, $\eta = 0.01$, the luminosity of the AGN expected from the inflow rate we measure would be $L_{\rm{AGN}} = \eta \, \dot{M}_{in} \, c^2 \sim 10^{46} \; \rm{erg \, s}^{-1} $, where $c$ is the speed of light. Assuming $L_{\rm{AGN}} = 100 \times L_{H\alpha}$ \citep{Heckman+04, Greene_Ho_2005}, we would then expect $L_{H\alpha} \sim 10^{44} \; \rm{erg \, s}^{-1}  $, that is almost 5 dex higher than the observed luminosity (Sect. \ref{sect:sfh}). Conversely, for the measured $L_{H\alpha} \sim 2 \times 10^{39} \; \rm{erg \, s}^{-1}$, we would expect $L_{\rm{AGN}} \sim 2 \times 10^{41} \; \rm{erg \, s}^{-1} $, implying $\dot{M}_{in} \sim 3 \times 10^{-4} M_\odot \rm{yr}^{-1}$ (assuming $\eta=0.01$), i.e. about 5 dex lower than the estimated inflowing mass. Thus, it seems unlikely that the infalling gas is powering the AGN. However, the intrinsic flux of the emission lines could be significantly underestimated due to orientation effects. Furthermore, the inflowing material could end up in the accretion disk of the AGN rather than directly into the black hole, so that the inflowing gas and the accretion rate might not be directly correlated. Therefore, with current data, we cannot rule out the AGN. Spatially resolved observations of the gas are needed to further probe this possibility.

\section{Conclusions}\label{sect:conclusions}

We have reported the spectroscopic detection of neutral gas inflow in a massive  quiescent galaxy, GLASS-180009, with $M_* \simeq 4 \times 10^{10} M_\odot$ observed at $z_{spec}=2.6576$ with JWST. From the absorption features of the NaI at $\lambda \lambda \, 5890, 5896 \AA$, we have derived the mass, $1.6^{+0.1}_{-0.1}\times 10^8 M_\odot$, and rate, $19^{+6}_{-7} \, M_\odot \, \rm{yr}^{-1}$, of the inflowing neutral gas. 

With current observations, we cannot determine the source of the neutral gas. By checking its environment, we found that GLASS-180009, which resides in an overdensity, is surrounded by several galaxies within a projected radius of 150 kpc at comparable redshifts, and they are potential source candidates providing the inflowing material. Other viable sources of the inflow are the IGM and cosmic filaments. Instead, a galactic fountain seems unlikely.

An inflow of neutral gas is usually associated with enhanced SF. However, we estimated an upper limit to the SFR $\leq 0.2 \, M_\odot \, \rm{yr}^{-1}$ and sSFR $\leq 0.5 \times 10^{-11} \, \rm{yr}^{-1} $ , implying that GLASS-180009 is quiescent. Furthermore, according to its SFH, the galaxy has remained passive during the past $\sim 1$ Gyr. On the basis of the estimated properties, we conclude that the inflowing gas is likely not massive or dense enough, or that the duration of the inflow is too short to enhance significant SF. Finally, it is also possible that the inflowing material is accreted at the center of the galaxy, fueling the AGN rather than the SF, although no strong AGN lines are detected.

The detection of neutral gas inflow in a passive galaxy offers a unique view of the post-quenching phase of quiescent galaxies at high redshifts. This system could prove that quiescent galaxies can accrete material without necessarily triggering the SF or AGN activity long after their quenching. The fact that this galaxy resides in an overdensity makes this case even more interesting. Understanding the origin and fate of the inflowing gas in this galaxy would provide a unique insight into the complexity of the gas-cycle of galaxies in dense environments, which is crucial for models of galaxy evolution. To this aim, targeted, deeper multiwavelength observations of this galaxy are necessary. Observations with the IFU of JWST would allow us to recover the spatially resolved kinematics of the neutral gas and trace its origin as well as its spatial distribution. Observations with higher S/N would allow us to constrain the properties of the gas using other features, such as MgII and FeI. ALMA observations targeting neutral and molecular hydrogen would provide more robust constraints on the properties of the inflowing gas.

\begin{acknowledgements}

The authors acknowledge financial support from programs HST-GO-17231 and HST-GO-17226, provided through a grant from the STScI under NASA contract NAS5-26555,  and program JWST-GO-02561, provided through a grant from the STScI under NASA contract NAS5-03127. F.L.B. acknowledges support from the INAF minigrant 1.05.23.04.01. S.B. is supported by ERC grant 101076080. P.J.W. acknowledges support from the INAF Large Grant 2022 `Extragalactic Surveys with JWST' (PI Pentericci). P.J.W. is supported by the European Union - NextGenerationEU RFF M4C2 1.1 PRIN 2022 project 2022ZSL4BL INSIGHT. P.J.W. acknowledges support from the INAF Mini Grant `1.05.24.07.01 RSN1: Spatially Resolved Near-IR Emission of Intermediate-Redshift Jellyfish Galaxies' (PI Watson).  X.W. is supported by the China Manned Space Program with grant no. CMS-CSST-2025-A06, the National Natural Science Foundation of China (grant 12373009), the CAS Project for Young Scientists in Basic Research Grant No. YSBR-062, the Fundamental Research Funds for the Central Universities, the Xiaomi Young Talents Program.

\textit{Software:} \texttt{NumPy} \citep{numpy}; \texttt{SciPy} \citep{scipy}; \texttt{Astropy} \citep{astropy}; \texttt{matplotlib} \citep{matplotlib}; \texttt{pPXF} \citep{ppxf_2023}; \texttt{VoigtFit} \citep{voigtfit}; \texttt{emcee} \citep{emcee} \texttt{BAGPIPES} \citep{bagpipes}; \texttt{Pathfinder} \citep{pathfinder}. 
\end{acknowledgements}

\appendix

\section{Wavelength calibration and photometric rescaling}\label{app:calibration}

\begin{figure*}
    \includegraphics[width=\textwidth]{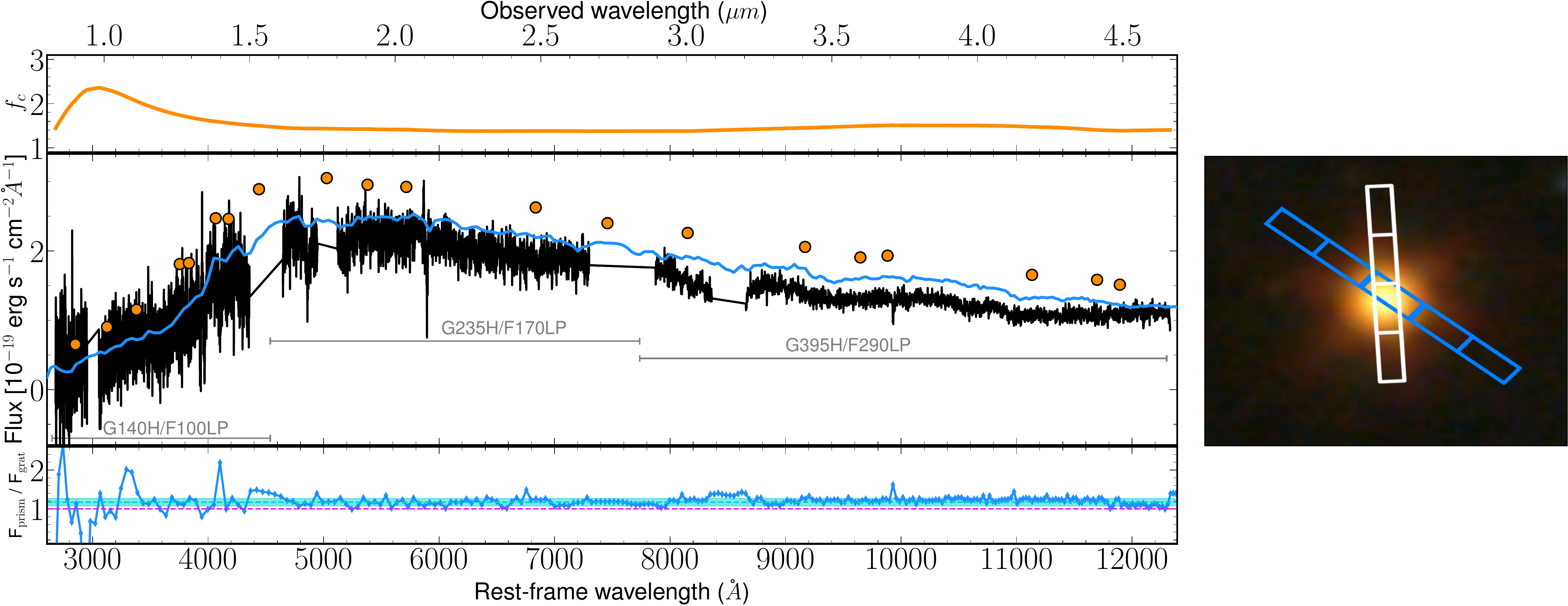}
    \caption{\textit{Left}: Spectroscopic and photometric data of GLASS-180009. In the central panel we show original (i.e. non scaled to the photometry) grating (black curve) and PRISM (blue curve) spectra; the orange circles are the photometric data. In the upper panel we show the correction factor, $f_c$, applied over the grating spectrum to rescale it to the photometric data. We estimate a median value $f_c = 1.5$, with no evident dependence on the wavelength. In the lower panel we show the flux ratio between the PRISM spectrum and the high-resolution spectrum with blue diamonds. The dashed blue lines and shaded region indicate the median and 1$\sigma$ error of the ratio. The dashed magenta horizontal line marks the 1-to-1 ratio. The flux ratio shows that there is no evident issue with the wavelength calibration. \textit{Right:} F444W image of the galaxy. The footprints of the NIRSpec microshutters for the gratings and PRISM are over-plotted in white and blue, respectively.}
    \label{fig:spec_original}
    
\end{figure*}

Concerning the reliability of the JWST spectra, \citet{Bevacqua+25} showed that the JWST spectra might suffer calibration issues \citep[see also][]{DR3}. In particular, the authors pointed out a mismatch between the fluxes of the PRISM and grating spectra that increases with the wavelength. For this reason, we checked the reliability of the wavelength calibration by comparing the fluxes of the PRISM and the grating spectra (degraded to the resolution of the PRISM), as shown in the lower panels of Fig. \ref{fig:spec_original}. We found that the flux of the grating spectra is about 15\% lower than that of the prism. However, we did not find evidence for a dependence on the wavelength, implying no evident calibration issues. 

We also compared the flux of the grating spectra with that of the photometric data. To this aim, we integrated the flux of the spectra over the NIRCam filters, weighted it with the transmission functions, and compared it with the NIRCam photometry. We found that the photometric fluxes are, on average, $\sim 50\%$ higher than those of the spectra, with a variation of about $\pm20\%$ (consistent with the uncertainties of the spectrum and photometry combined), but without any significant dependence on the wavelength. Therefore, we rescaled the flux of the high-resolution spectrum to the photometry, using a polynomial of degree 8.

\section{Estimates of effective radius and S\'{e}rsic index}\label{app:pysersic}

\begin{figure*}
    \centering
    \includegraphics[width=\textwidth]{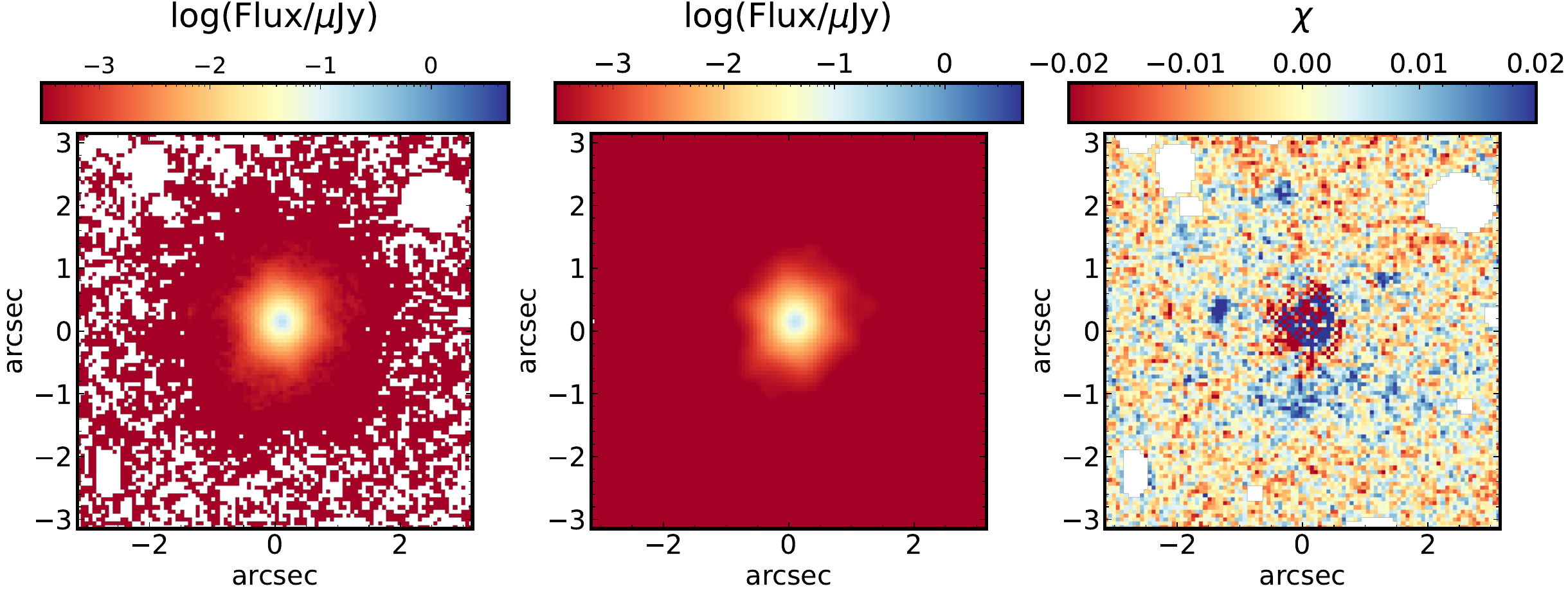}
    \caption{Fit of the F277W image with \texttt{pysersic}. The left and central panels are the observed and best-fit model images, while the right panel shows the $\chi$ map, i.e. the residuals divided by the errors.}
    \label{fig:pysersic}
\end{figure*}

To estimate the effective radius and S\'{e}rsic index, we use \texttt{pysersic} \citep{pysersic}, a Bayesian inference fitting tool, to fit the F277W image from UNCOVER with a 1D S\'{e}rsic profile. Background sources have been masked using the UNCOVER DR3 segmentation map. We sample the marginalized posteriors using a no U-turn sampler using two chains with 750 warm-up and 500 sampling steps each \citep{Hoffman14}. We show the \texttt{pysersic} best-fitting image in Fig \ref{fig:pysersic} . As a result, we found best-fitting $R_e =1.05^{+0.01}_{-0.01}$ kpc and $n = 2.15^{+0.01}_{-0.01}$.

\section{Testing the fits of the stellar population properties}\label{app:test_sps}

\begin{figure}
    \centering
    \includegraphics[width=\columnwidth]{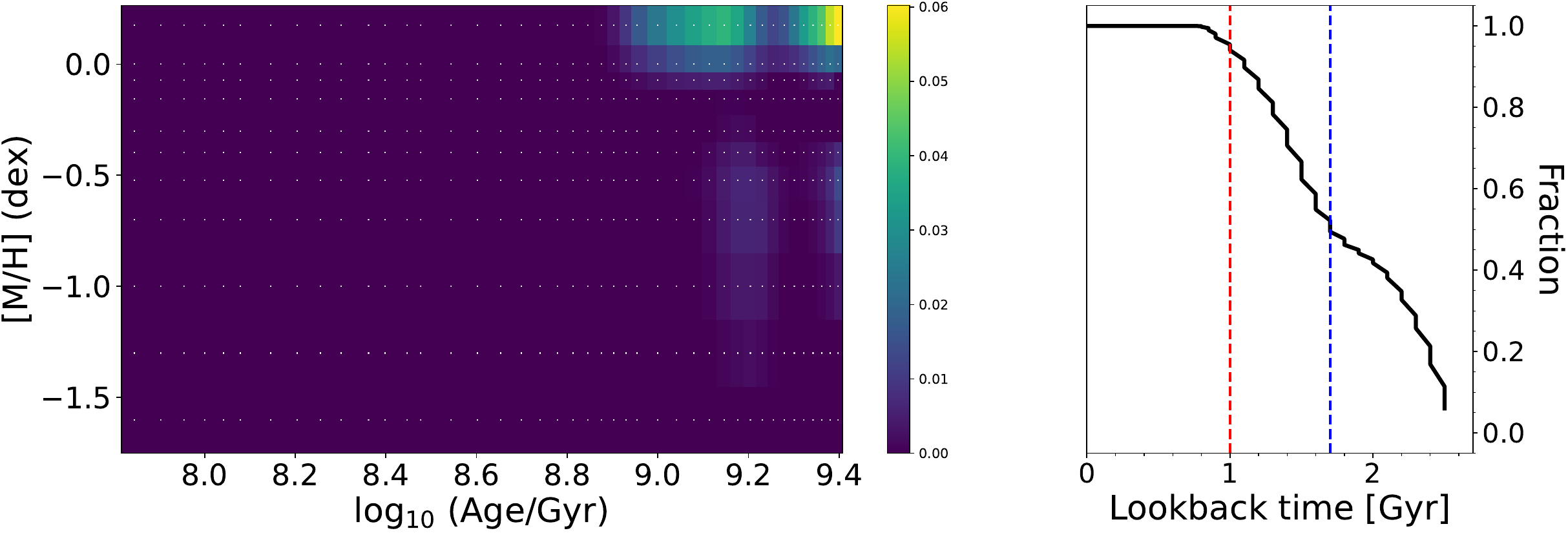}
    \caption{Star formation history derived with \texttt{pPXF}, as in Fig. \ref{fig:ppxf_bf}, but using the BC16 models.}
    \label{fig:sfh_ppxf_bc16}
\end{figure}

In this section, we test the reliability of the derived stellar population properties. In particular, we want to probe whether including very young models affects the results. To this aim, we considered the stellar population models of \citet[][updated to version 2016; hereafter, BC16]{BC03}. Following the same methods described in Sect. \ref{sect:spmodeling}, we fit the high-resolution JWST spectrum with \texttt{pPXF} adopting the BC16 models and including ages as young as 1 Myr. However, in this case, we limit the fit to the rest frame $\lambda < 1 \mu m$, because the BC16 models have a lower resolution than the JWST spectrum at longer wavelengths. From this fit, we estimate a mass-weighted age of 1.4 Gyr and a supersolar metallicity, [M/H] = 0.3 dex, approximately 0.4 Gyr lower and 0.6 dex higher than the values estimated with the EMILES models. However, as evident from the SFH shown in Fig. \ref{fig:sfh_ppxf_bc16}, we find that no young components contribute to the best-fitting spectrum. Moreover, we estimate the same quenching time $t_{95} \simeq 1$ Gyr as with the EMILES models. We verified that the difference in the results is due to both the different wavelength range fit and intrinsic differences in the adopted models (isochrones, stellar libraries, etc.).

\startlongtable
\begin{deluxetable*}{lllll}\label{tab:bagpipes_input}
\tablewidth{0pt}
\tablecaption{\texttt{Bagpipes} input parameters}
\tablehead{
\colhead{Component} & \colhead{Parameter} & \colhead{Symbol/Unit} & \colhead{Range} & \colhead{Prior}
}
\startdata
General & Redshift & $z$ & $(2.5, \, 2.8)$ & Uniform \\
  & Stellar velocity dispersion & $\sigma_*$/ km s$^{-1}$ & (50, 500) & Logarithmic \\
\hline
SFH & Total stellar mass formed & $M_{*, tot}/M_\odot$ & $(1, 10^{13})$ & Logarithmic \\
 & Double-power-law falling slope & $\alpha$ & $(0.1, 1000)$ & Logarithmic \\
  & Double-power-law rising slope & $\beta$ & $(0.1, 1000)$ & Logarithmic \\
  & \footnotesize{Double-power-law turnover time} & $\tau_{\rm{dpl}}$ & $(0.1, \, 2.5)$ & Uniform \\
\hline 
Dust & V-Band attenuation & $A_V$/mag & $(0, 4)$ & Uniform \\
\hline
Nebular & Ionization parameter & U & $(10^{-4}, 10^{-2})$ & Logarithmic\\
 & Gas metallicity & $Z_g/Z_\odot$ & $(0.00355, 3.55)$ & Logarithmic \\
\hline
Noise & White noise scaling & $s$ & $(0.1, 10)$ & Logarithmic \\
\enddata
\tablecomments{For the dust, we assume a \citet{Calzetti00} attenuation law.}
\end{deluxetable*}

As a further independent check, we fit the high-resolution spectrum with \texttt{Bagpipes} \citep{bagpipes}, adopting the same assumptions as for the fits with \texttt{pPXF} (Sect. \ref{sect:spmodeling}) and using the BC16 models. The only significant differences are that we parametrize the SFH with a double-power-law, we fit the emission lines with a nebular component, and we fit a white noise scaling factor. The summary of the input parameters is reported in Table \ref{tab:bagpipes_input}.

\begin{figure}
    \centering
    \includegraphics[width=\linewidth]{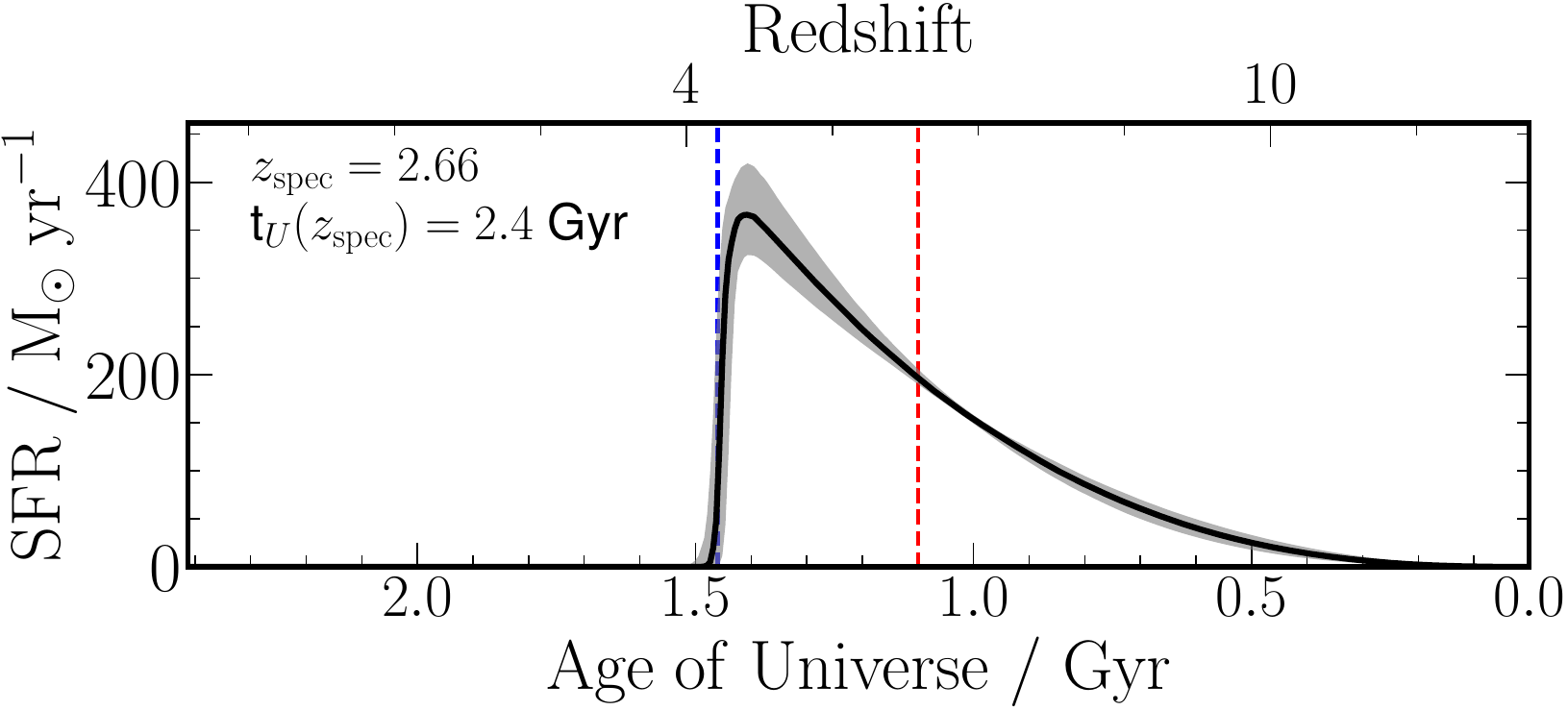}
    \caption{Star formation history derived with \texttt{Bagpipes} using the BC16 models. The red and blue lines indicate t$_{50}$ and t$_{90}$, respectively.}
    \label{fig:sfh_bagpipes}
\end{figure}

\begin{figure*}
    \centering
    \includegraphics[width=\textwidth]{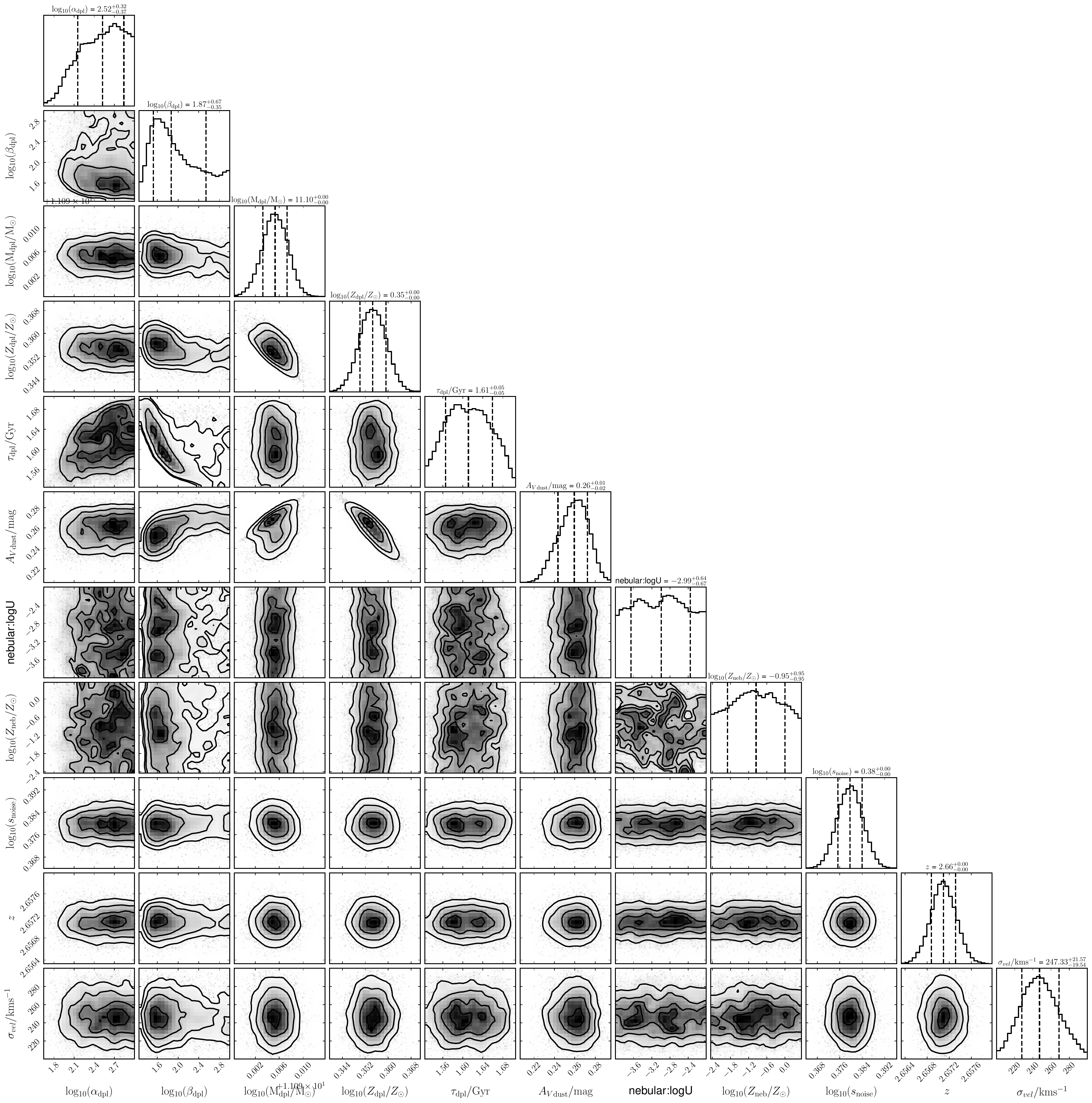}
    \caption{Full posterior distributions of the fitted parameters reported in Table \ref{tab:bagpipes_input}. Note that no correction is here applied for the magnification \citep{Marchesini+23}.}
    \label{fig:bagpipes_full_posterior}
\end{figure*}

We show the SFH derived with \texttt{Bagpipes} in Fig. \ref{fig:sfh_bagpipes}. The relevant stellar population properties are reported in Table \ref{tab:bagpipes_output}. The complete posterior distribution is shown in Fig. \ref{fig:bagpipes_full_posterior}. In general, we find consistent results with the fits of \texttt{pPXF} when fed with the BC16 models. 

We then conclude that the stellar age and metallicity depend on the adopted models, as well as on the wavelength range fitted \citep[see also ][]{Bevacqua+25}. However, investigating these differences is beyond the scope of this article. The main results of this section are that including very young models does not change the results and that all fits consistently indicate that the galaxy quenched $\sim 1$ Gyr before the observations without experiencing any other significant SF event afterwards.

\begin{deluxetable*}{c|c}\label{tab:bagpipes_output}
\tablewidth{0pt}
\tablecaption{Relevant stellar population parameters derived with \texttt{Bagpipes}}
\tablehead{
\colhead{Parameter} & \colhead{Value}
}
\startdata
$\log_{10}(M_*^{\rm{corr}}/M_\odot)$ & $10.66^{+0.02}_{-0.02}$\\
$\sigma_*$ / km s$^{-1}$ & $247^{+22}_{-20}$\\
$A_V$/mag & $0.26^{+0.01}_{-0.02}$\\
Age/Gyr & $1.32^{+0.06}_{-0.11}$ \\
$\log_{10}(Z_*/Z_\odot)$/ dex & $0.35^{+0.01}_{-0.01}$\\
SFR / $M_\odot$ yr$^{-1}$ & 0.00$^{+0.00}_{-0.00}$\\
$t_{50}$/Gyr & $1.40^{+0.08}_{-0.12}$\\
$t_{95}$/Gyr & $0.94^{+0.04}_{-0.03}$\\
\enddata
\tablecomments{$M_*^{\rm{corr}}$ is the mass of living stars, corrected for the magnification factor \citep[see ][]{Marchesini+23}. }
\end{deluxetable*}

\section{Estimate of the [Mg/Fe] abundance}\label{app:mgfe}

\begin{figure*}
    \includegraphics[width=0.32\textwidth]{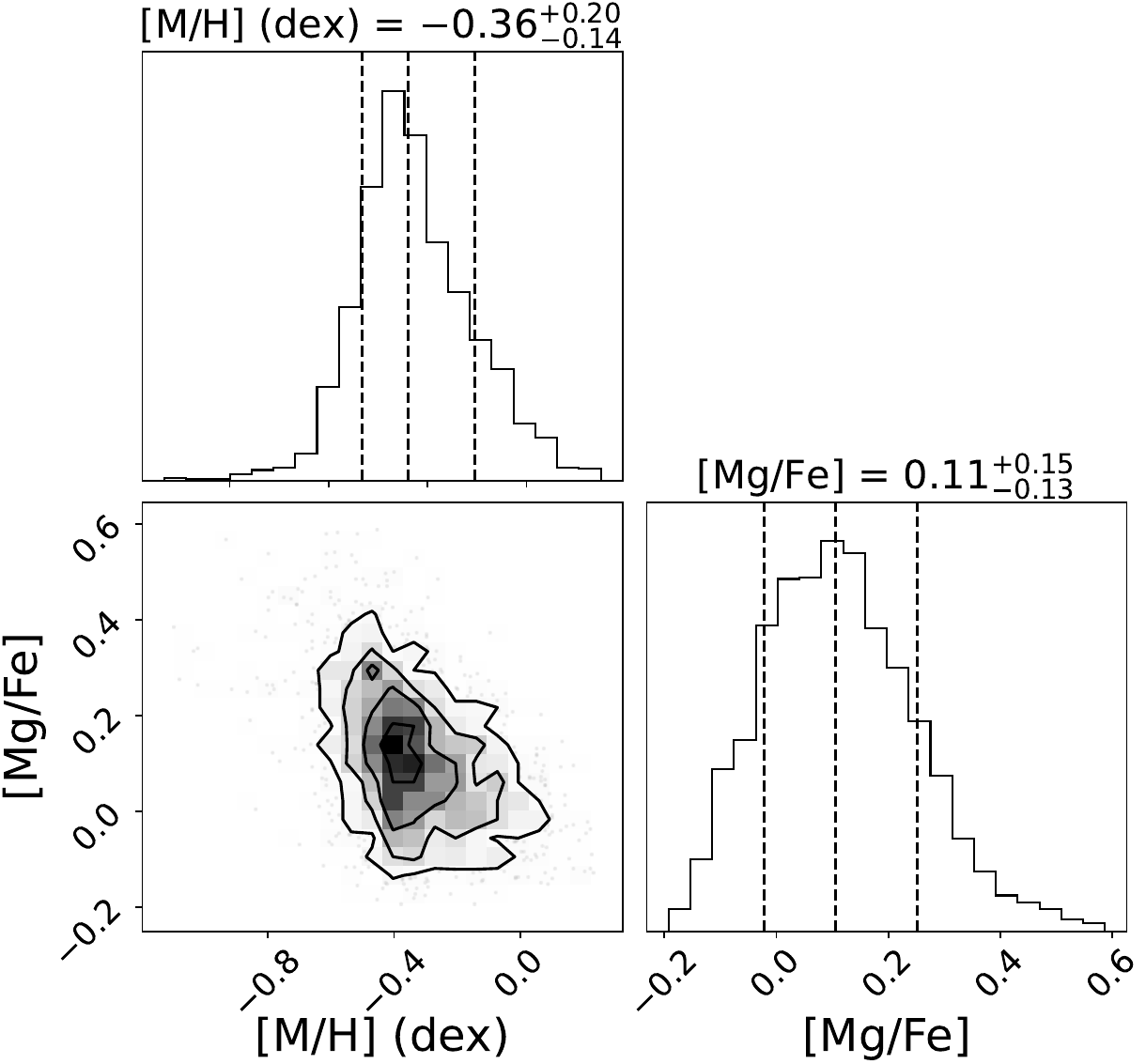}
    \includegraphics[width=0.32\textwidth]{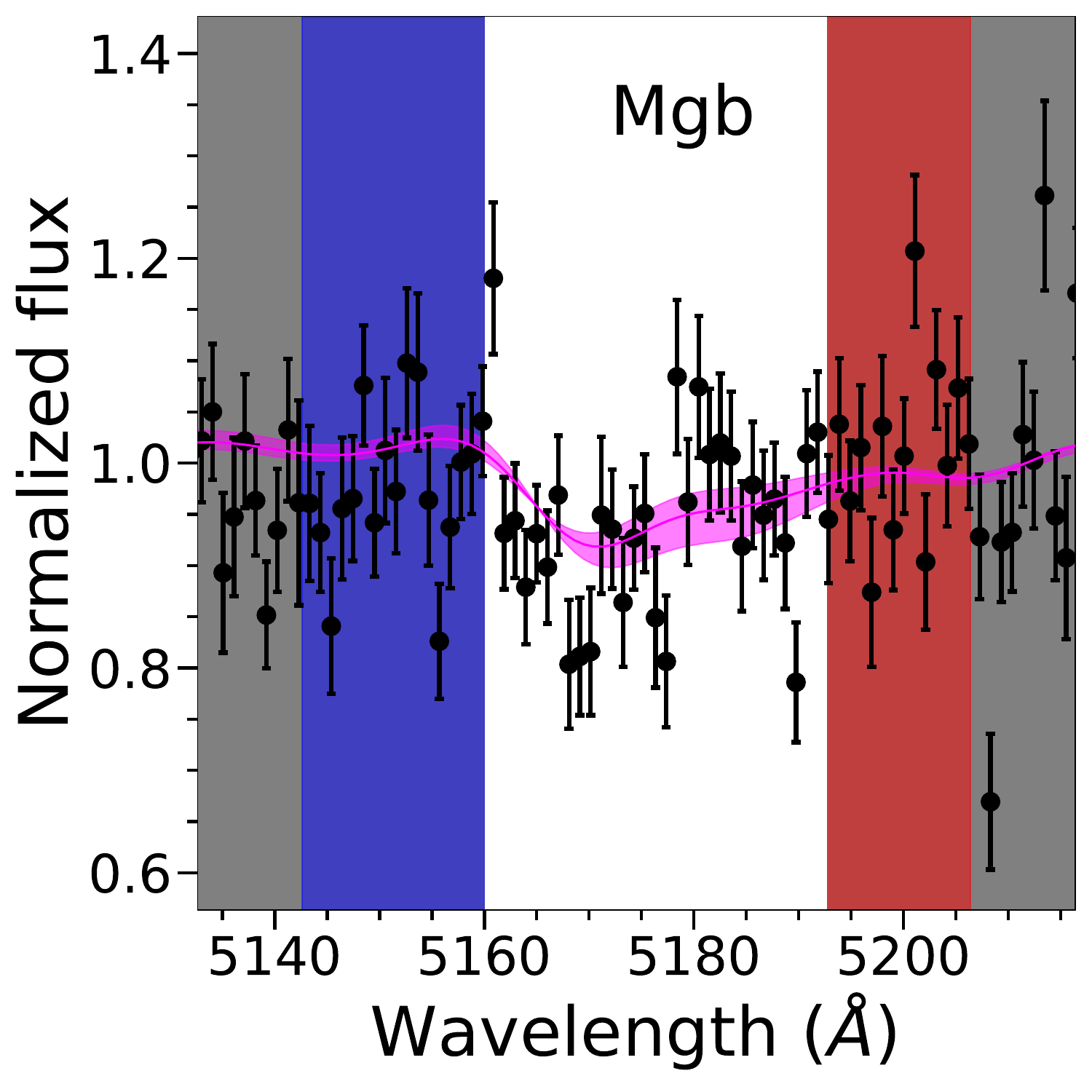}
    \includegraphics[width=0.32\textwidth]{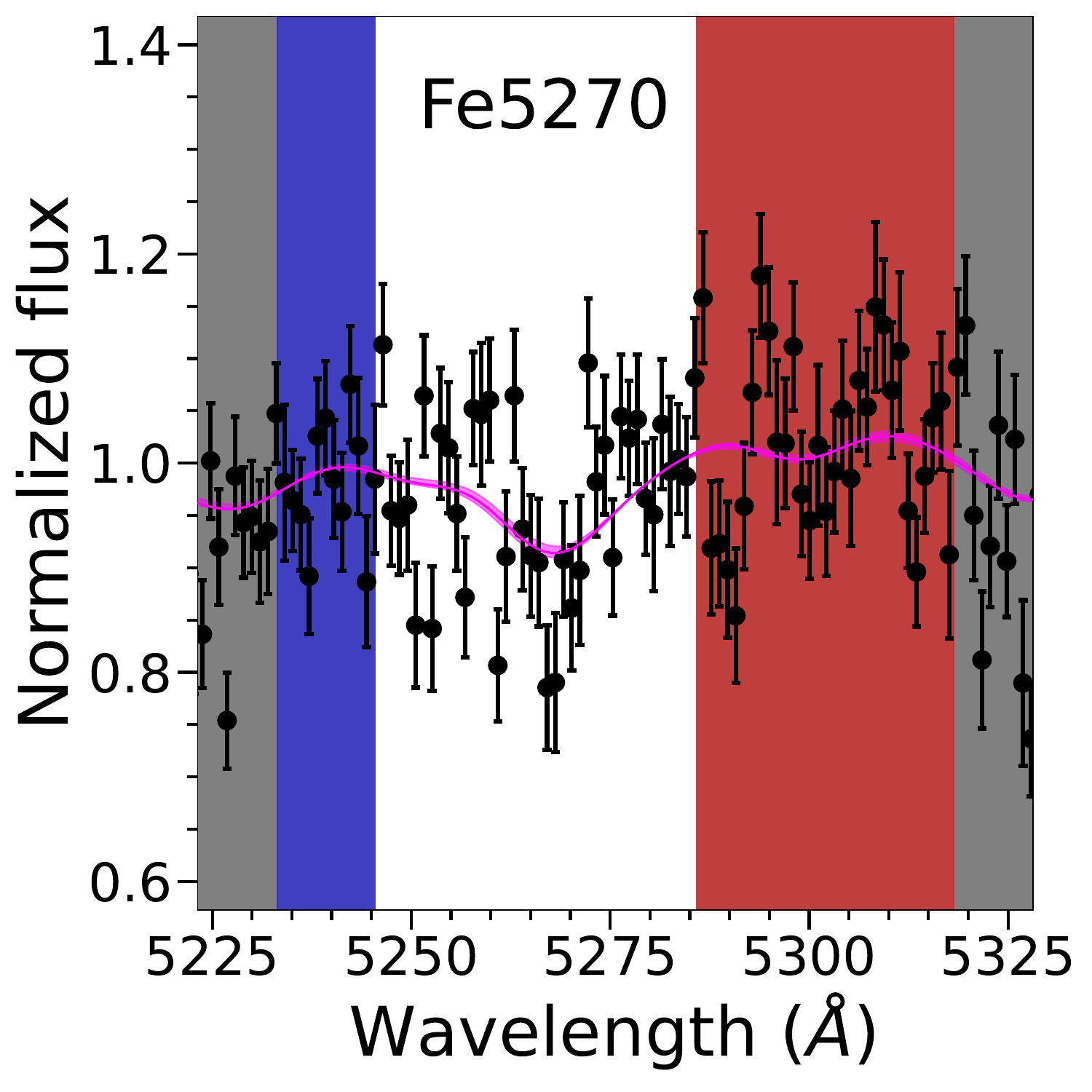}
    \caption{Best-fitting solution for the [Mg/Fe] abundance with the FIF method. In the left panel we show the corner plot of the posterior distributions. In the central and right panel we show the observed spectral indices (black circles) compared to the best-fitting models (magenta line). The blue and red vertical regions indicate the wavelength region of the blue and red pseudo-continua, respectively, while the white central region is the feature.}
    \label{fig:mgfe}
\end{figure*}

We estimate the [Mg/Fe] abundance of GLASS-180009 by fitting the spectral indices Mgb and Fe5270. The Mgb is considered as a tracer of the Mg-abundance, while Fe5270 as a tracer of the iron abundance \citep[see, e.g.,][]{Thomas+03}. Other reliable tracers of the iron abundance are Fe4383 and Fe5335. However, the former falls into the detector gap of the NIRSpec spectra, and the latter is significantly affected by bad pixels, so we decided not to use it.

We estimate [Mg/Fe] with the Bayesian Full Index Fitting (FIF) method described in \citet{FIF1, FIF2}. The FIF method performs a pixel-by-pixel fit of all the spectral pixels within the band passes of the fitted spectral features. Here, we adopt \texttt{emcee} to maximize the following likelihood function:
\begin{equation}
\ln(\textbf{O}|\textbf{S}) = - \frac{1}{2}\sum_{n}\left[\frac{(O_n-M_n)^2}{\sigma_n^2} - \ln\left( \frac{1}{\sigma^2_n}\right) \right] \; ,
\end{equation}
where the sum extends over all the spectral pixels within the band passes fitted, with $O_n$ and $M_n$ being the observed and the model fluxes, respectively, of the $n$-th spectral pixel, while $\sigma_n$ is the uncertainty on $O_n$. We perform the fit by keeping the age fixed\footnote{This is because the robust spectral indices sensitive to the age are the Balmer lines, which are however affected by line emissions.} to the light-weighted best-fitting age of \texttt{pPXF}, 1.65 Gyr, while varying the metallicity and [$\alpha$/Fe]. As for the templates, we adopted the semi-empirical MILES (sMILES) models with variable $\alpha$-enhancement \citep{smiles}. 

\begin{figure}
    \centering
    \includegraphics[width=\columnwidth]{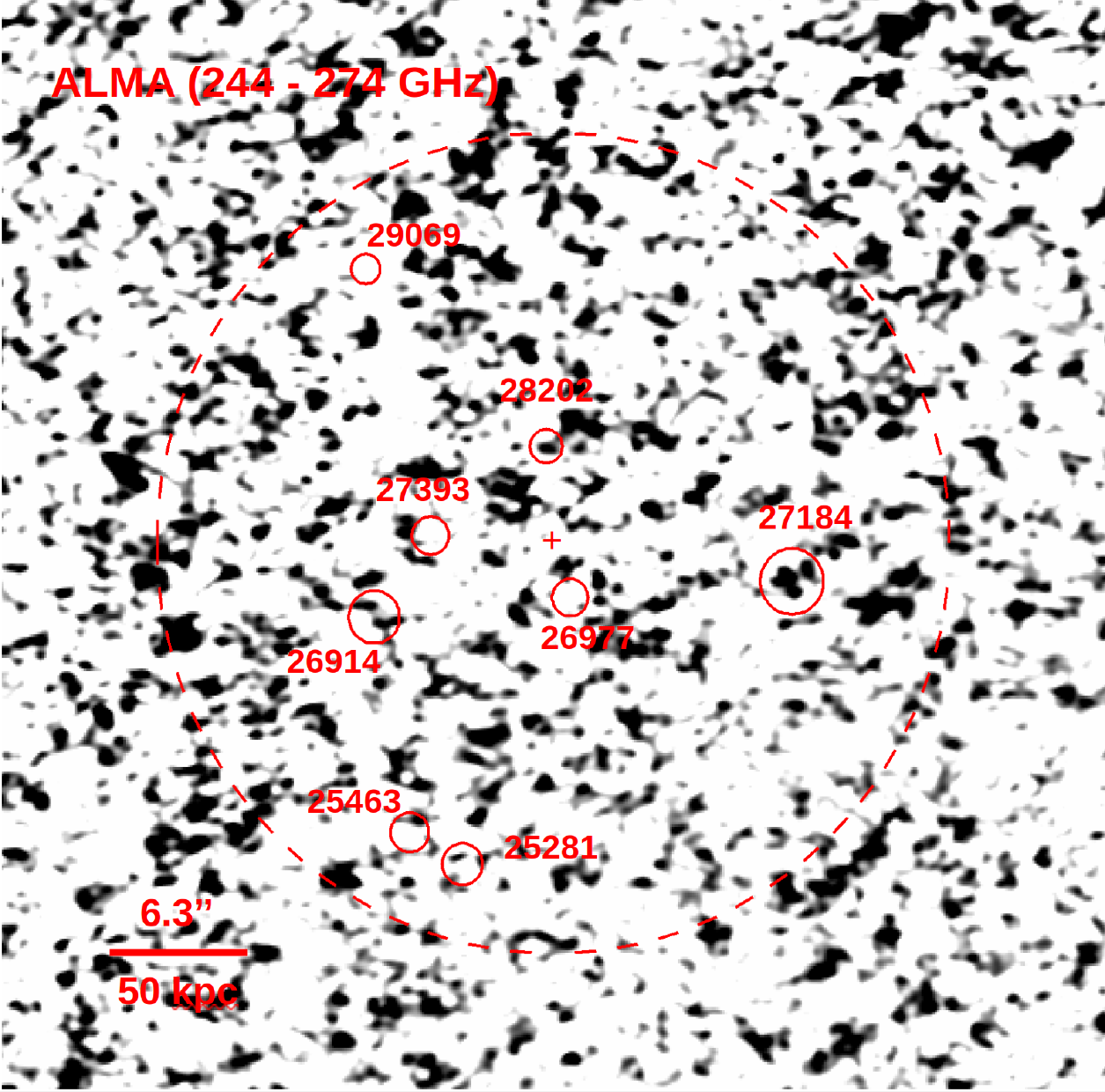}
    \caption{ALMA observations of the same sky region as in Fig. \ref{fig:environment}; the symbols are the same.}
    \label{fig:alma-dualz}
\end{figure}

From this fit, we estimate [M/H] = $-0.36^{+0.20}_{-0.14}$ dex (consistent with the full-spectral fitting) and $\rm{[Mg/Fe]} = 0.11^{+0.15}_{-0.13}$ dex. We show the posterior distribution and the best fits in Fig. \ref{fig:mgfe}. We mention that varying the age within the errors provides consistent results. Overall, the quality of the spectrum is not sufficient to robustly constrain the [Mg/Fe], but our fits indicate a moderate $\alpha$-enhancement, comparable to quiescent galaxies at lower redshifts (e.g., \citealt{Bevacqua+23}).

\section{ALMA-DUALZ observations}\label{app:alma}

In Fig. \ref{fig:alma-dualz}, we show the ALMA imaging from DUALZ \citep{dualz}. The image shows the same sky region shown in Fig. \ref{fig:environment} in the frequency band 244-274 GHz, corresponding to rest frame $\lambda \sim 0.3 \; mm$. The image is dominated by the noise and does not show any clear detection of gas around GLASS-180009 or the surrounding galaxies. Therefore, from these observations, we cannot probe the origin and distribution of the neutral gas.

\bibliography{biblio}{}
\bibliographystyle{aasjournalv7}



\end{document}